\documentclass[sigplan,nonacm]{acmart}

\PassOptionsToPackage{table}{xcolor}
\usepackage[ruled,vlined,linesnumbered]{algorithm2e}
\usepackage{enumitem}
\usepackage{amsmath}
\usepackage{graphicx}
\usepackage{booktabs}
\usepackage{multirow}
\usepackage{array}
\usepackage[table]{xcolor}
\usepackage{tabularx}
\usepackage[symbol]{footmisc}
\usepackage{pifont}
\usepackage{subcaption}
\usepackage{caption}
\usepackage{siunitx}
\usepackage{tikz}

\AtBeginDocument{%
  }

\begin{document}

\title{Multipath Memory Access: Breaking Host--GPU Bandwidth Bottlenecks in LLM Serving}

\author{Lingfeng Tang\textsuperscript{\textdagger,1} Daoping Zhang\textsuperscript{\textdagger,1} Junjie Chen\textsuperscript{1} Peihao Huang\textsuperscript{*,1} Feng Jin\textsuperscript{2} Chengguang Xu\textsuperscript{2} Yuxin Chen\textsuperscript{2} Feiqiang Sun\textsuperscript{2} Guo Chen\textsuperscript{*,1}}
\affiliation{%
  \institution{\vspace{12pt}\textsuperscript{1}Hunan University \quad \textsuperscript{2}Tencent}
  \country{}}
\renewcommand{\shortauthors}{Tang, Zhang, et al.}

\begin{abstract}
Host--GPU data movement has become a latency-critical bottleneck in LLM serving, surfacing in common paths such as model-weight movement and KV cache offload/fetch. Today, each host--GPU copy is effectively confined to the PCIe path of the target GPU, even though modern multi-GPU servers contain additional PCIe links on peer GPUs and high-bandwidth GPU interconnects. This leaves substantial intra-server I/O capacity unused. To address this issue, we present Multipath Memory Access (MMA), a software-defined multipath memory access system for host--GPU data transfer. To the best of our knowledge, MMA is the first software-defined system to enable efficient multipath host--GPU data transfer within a single multi-GPU server. MMA expands a single host--GPU copy across available direct and relay paths without hardware, driver, or application changes. It preserves CUDA stream semantics through a dependency-preserving Dummy Task, coordinates distributed micro-transfer completion through a lightweight synchronization mechanism, and uses queue backpressure to route traffic without explicit link-state feedback. On an 8-GPU NVIDIA H20 server, MMA achieves 245\,GB/s peak host-to-GPU bandwidth, a 4.62$\times$ improvement over native CUDA copies, and reduces TTFT for KV cache fetching by 1.14--2.38$\times$ and model wake-up/switching latency by 1.12--2.48$\times$.
\end{abstract}

\maketitle
\begingroup
\renewcommand{\thefootnote}{\fnsymbol{footnote}}
\footnotetext[2]{Equal contribution (co-first authors).}
\footnotetext[1]{Corresponding author.}
\endgroup

%% ============================================================
%% SECTION 1: Introduction
%% ============================================================
\section{Introduction}
\label{sec:intro}

\begin{figure}[t]
\centering
\includegraphics[width=\columnwidth]{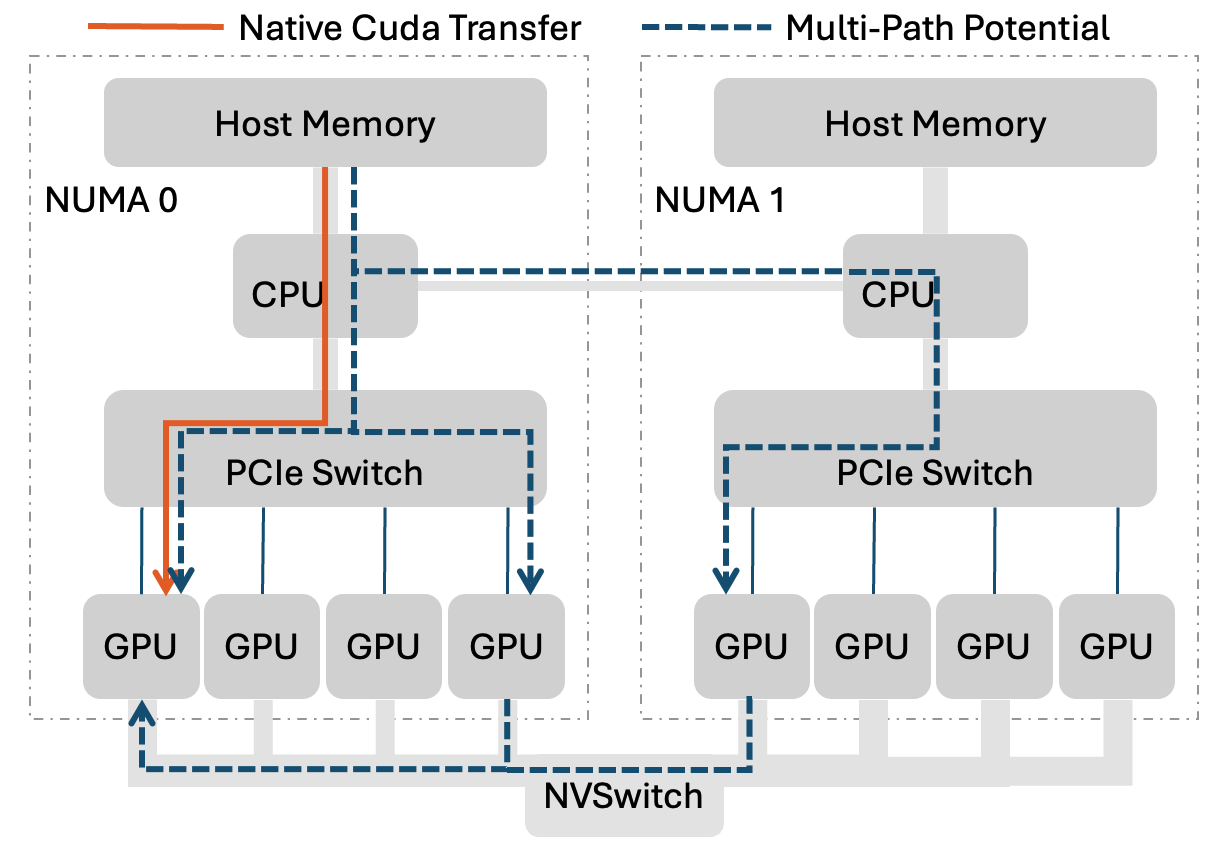}
\vspace{-10pt}
\caption{Simplified intra-server PCIe topology. Each NUMA node contains two PCIe switches; eight GPUs are interconnected via NVLink.}
\vspace{-10pt}
\label{fig:topology}
\end{figure}

\begin{figure*}[htbp]
    \centering
    \begin{minipage}{0.46\textwidth}
        \centering
        \captionsetup{skip=5pt}
        \includegraphics[width=\linewidth]{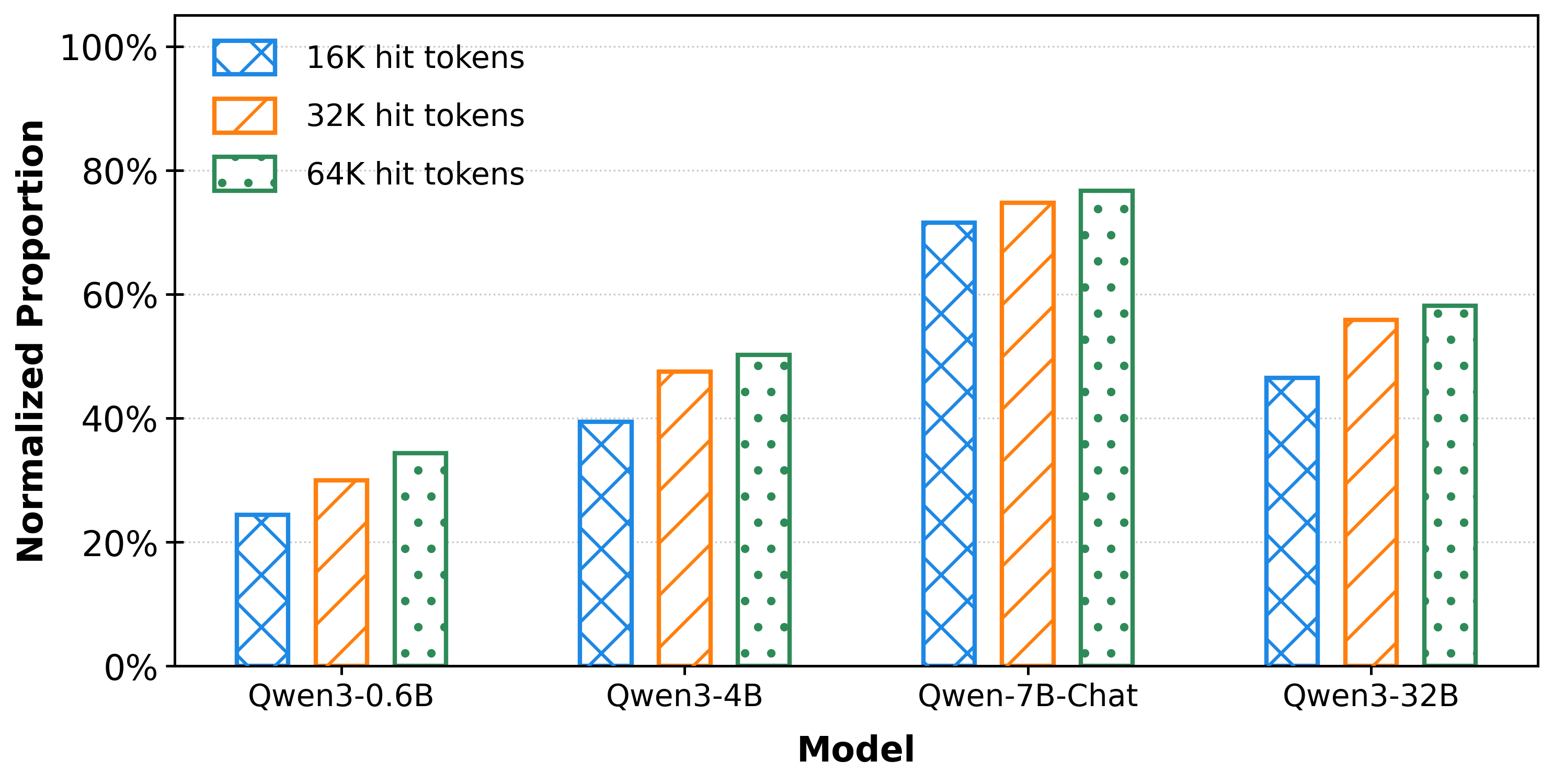}
        \captionof{figure}{The proportion of prefix-cache fetching time in TTFT under different hit-token lengths and models}
        \label{fig:h2d}
    \end{minipage}
    \hfill
    \begin{minipage}{0.46\textwidth}
        \centering
        \captionsetup{skip=5pt}
        \includegraphics[width=\linewidth]{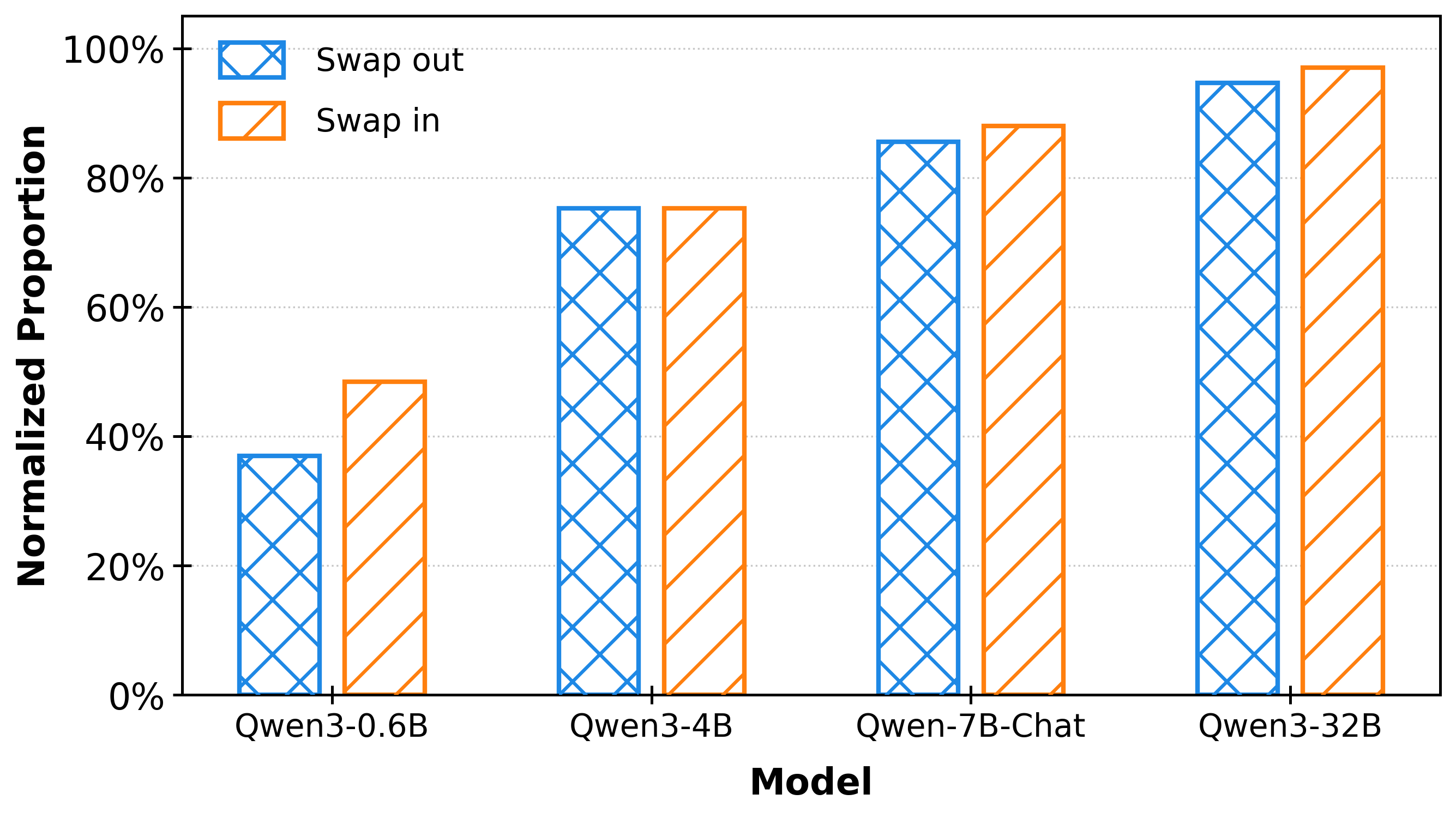}
        \captionof{figure}{The proportion of H2D/D2H transfer time in swap-in and swap-out latency under different models}
        \label{fig:sleep03}
    \end{minipage}
    \vspace{-10pt}
\end{figure*}

Host--GPU data movement is now a latency-critical component of large language model (LLM) serving~\cite{crucial2,crucial1}.
This movement appears in several serving paths; two common examples are:
\emph{model-weight movement}---loading or reloading parameter tensors between host memory and GPU memory during model wake-up, switching, or checkpoint restore~\cite{ServerlessLLM,pipeswitch,crucial3,switch1};
and \emph{KV cache transfers}---offloading cached key--value tensors to host memory and fetching them back to skip redundant prefill computation~\cite{gao2024cost,gim2024prompt,pan2024marconi,yanglearned,zheng2024sglang}.
In today's systems, however, each individual host--GPU copy is bound to the target GPU's PCIe path, making a single PCIe link the bottleneck for that copy even when other links in the server are idle or lightly loaded.
On our 8$\times$H20 testbed, the bottleneck is severe: switching a 32\,B-parameter model takes $\sim$2.5 seconds even on PCIe~5.0---well above typical sub-second TTFT targets (\S\ref{Motivation}, Figure~\ref{fig:sleep03}), and loading a 64\,k-token KV cache consumes up to 70\% of time-to-first-token (Figure~\ref{fig:h2d}).

Yet the server contains much larger latent aggregate I/O capacity than a single PCIe link can deliver (Figure~\ref{fig:topology}).
Seven \emph{peer} GPUs each have their own PCIe link to host memory and a high-bandwidth NVLink path to the target GPU.
If a peer GPU reads data from host DRAM over its local PCIe link and forwards it to the target via NVLink, it effectively acts as a \emph{relay node}, opening an additional host-to-GPU data path.
In the upper-bound case where all seven peer PCIe links are available for relay, the theoretical host-read bandwidth rises from 64\,GB/s to $8 \times 64 = 512$\,GB/s---an $8\times$ increase---at which point host DRAM bandwidth and the inter-socket interconnect, not PCIe, become the binding constraints.
This gap is architectural: today's GPU runtime abstraction exposes the target GPU's PCIe path as the only data path for a host--GPU copy, leaving peer PCIe links and the GPU interconnect outside the transfer abstraction.
CUDA's memory-copy interface, as the dominant instance of this abstraction, provides no way to recruit these peer paths even when they have spare capacity.

Exploiting this capacity, however, is non-trivial.
Three properties of the GPU execution model stand in the way.
\textbf{(C1)}~CUDA's asynchronous execution model binds each transfer to a fixed path at enqueue time and provides no mechanism to reroute it after submission; any interception must occur before hardware dispatch while preserving stream ordering and event dependencies (\S\ref{sec:Opportunity and Challenges}).
\textbf{(C2)}~CUDA streams expose each memory copy as a single ordered task whose completion releases downstream work. Multipath transfer, however, completes only after data fragments across multiple direct and relay paths have all arrived; this distributed completion must be represented as a single stream-visible event without breaking CUDA's dependency semantics.
\textbf{(C3)}~Unlike network fabrics with explicit congestion signals, PCIe exposes no feedback to software, and the relay paths span heterogeneous links with widely varying bandwidths; without per-link congestion awareness, any static splitting strategy inevitably creates stragglers.

We present \textbf{MMA (Multipath Memory Access)}, a software-defined multipath system for bidirectional host--GPU data movement.
MMA's key insight is that peer GPUs can act as software-defined relay nodes: their independent PCIe links and the GPU interconnect together form a multipath fabric for host--GPU memory copies, without hardware modification.
Three mechanisms realize this insight:
a \emph{Transfer Task Interceptor} that records intercepted copies as Transfer Tasks and, for stream-ordered copies, uses a dependency-preserving placeholder to defer path selection before CUDA binds the copy to a fixed PCIe path~(C1);
a \emph{spin-kernel synchronization} mechanism that aggregates distributed micro-transfer completion into one stream-visible completion event~(C2);
and a \emph{pull-based path selector} that infers per-link congestion from buffer occupancy and balances traffic without explicit feedback~(C3).
MMA is implemented as a ${\sim}$3\,K-line C++ shared library~\cite{MMA-OPENS} activated via \texttt{LD\_PRELOAD}; existing CUDA applications---including vLLM and LMCache---benefit with zero code changes.

We evaluate MMA on an 8$\times$NVIDIA H20 server (PCIe 5.0 $\times$16, NVLink 4.0, dual-socket AMD EPYC 9654):

\begin{itemize}[leftmargin=*,nosep]
\item \textbf{Microbenchmark:} MMA achieves 245\,GB/s peak H2D bandwidth---\textbf{4.62$\times$} over the 53\,GB/s single-PCIe baseline.
Bandwidth saturates at six relay GPUs, where the inter-socket xGMI3 link becomes the residual bottleneck.
\item \textbf{End-to-end:} On LMCache+vLLM serving workloads, MMA reduces TTFT for KV cache loading by 1.14--2.38$\times$ and model-switching latency by 1.12--2.48$\times$.
\end{itemize}

\noindent\textbf{Contributions.}
\begin{enumerate}[leftmargin=*,nosep]
\item \textbf{Multipath memory access abstraction.}
To the best of our knowledge, MMA is the first software-defined multipath system for host--GPU memory copies within a single server. MMA expands a single CUDA memory copy beyond the target GPU's PCIe path by recruiting peer GPUs as relay nodes through their PCIe links and the GPU interconnect, without requiring hardware, driver, or application changes.
\item \textbf{CUDA-transparent multipath design.}
We design three mechanisms that make multipath host--GPU transfer compatible with CUDA's execution semantics: a Transfer Task Interceptor that defers path selection before CUDA binds a copy to a fixed PCIe path; a spin-kernel synchronization mechanism that aggregates distributed micro-transfer completion into one stream-visible completion event; and a pull-based path selector that balances traffic across heterogeneous paths without explicit link-state feedback.
\item \textbf{Implementation and evaluation.}
We implement MMA as a ${\sim}$3\,K-line CUDA~12.8 user-space library deployable via \texttt{LD\_PRELOAD} and Python/C++ APIs. On an 8$\times$NVIDIA H20 server, MMA achieves 245\,GB/s peak host-to-GPU bandwidth, a 4.62$\times$ improvement over native CUDA copies, and reduces LLM serving TTFT and model-switching latency by 1.14--2.38$\times$ and 1.12--2.48$\times$, respectively.
\end{enumerate}

%% ============================================================
%% SECTION 2: Background and Motivation
%% ============================================================
\section{Background and Motivation}

\begin{table*}[t]
\centering
\caption{Bandwidth of key interconnects in contemporary multi-GPU servers. All values are unidirectional except host DRAM (aggregate read+write). Typical measured values are from published benchmarks and our testbed.}
\label{tab:performance}
\begin{tabular}{llcc}
\toprule
Interconnect & Generation & Theoretical BW & Typical Measured \\
\midrule
PCIe ($\times$16) & Gen 4.0~\cite{PCI-SIG_PCIe_5.0} & 32 GB/s & 24--26 GB/s \\
                   & Gen 5.0~\cite{PCI-SIG_PCIe_5.0} & 64 GB/s & 52--60 GB/s \\
Host DRAM          & DDR5-4800 (12\,ch\,/\,24\,ch) & 461\,/\,922 GB/s & 350--400\,/\,600--700 GB/s \\
NVLink (per GPU)   & 4.0 (18 links)~\cite{nvlink_4.0} & 450 GB/s & 400--478 GB/s \\
CPU interconnect$^\dagger$   & AMD xGMI3 (4 links)~\cite{epyc_9654} & ${\sim}$256 GB/s raw & --- \\
\bottomrule
\end{tabular}

\vspace{2pt}
{\small $^\dagger$CPU interconnect is not directly on the host--GPU transfer path but affects cross-NUMA access latency. The xGMI3 value is a raw one-way estimate for four 32\,Gbps-per-lane links before protocol overheads.}
\vspace{-10pt}
\end{table*}

LLM serving systems increasingly treat host memory as an extension of GPU memory.
When model weights, KV caches, or runtime states exceed GPU memory capacity or are temporarily inactive, serving frameworks offload them to host DRAM and later fetch them back on demand~\cite{aminabadi2022deepspeed,kwon2023efficient,mlpoffload,rajbhandari2021zero}.
This design enables larger models, longer contexts, and higher GPU utilization, but it also turns host--GPU data movement into a latency-critical part of serving.

This pressure appears in several serving paths; we focus on two representative ones.
First, KV cache offloading and fetching move previously computed key--value tensors between GPU memory and host memory, allowing repeated or long-context requests to skip redundant prefill computation.
Second, model-weight movement occurs during cold starts, model switching, checkpoint restore, or sleep/wake mechanisms, where large parameter tensors are moved between host memory and GPUs.
In both cases, each H2D or D2H copy is bound to the target GPU's PCIe path, making per-copy PCIe bandwidth the immediate bottleneck.

\subsection{Host--GPU transfers are latency-critical in LLM serving} \label{Motivation}

We quantify the performance impact of host--GPU data movement on two representative operations that exercise common host--GPU data-movement paths: KV cache movement and model-weight movement.
All measurements use our 8$\times$H20 testbed (hardware details in \S\ref{sec:design-eval}).

\textbf{Fetching cached KV pages from host memory accounts for up to 70\% of TTFT.}
Systems like vLLM~\cite{kwon2023efficient} and SGLang~\cite{zheng2024sglang} offload long-sequence KV caches to host memory when they cannot remain resident in GPU memory.
When a cached prefix is reused, the serving engine must fetch the corresponding KV pages from host memory back to the GPU before decoding can proceed.
This fetch latency directly affects time-to-first-token (TTFT), a latency-critical serving metric.
We measure TTFT for prefix-cache hits under different context lengths using the LMCache+vLLM~\cite{lmcache} inference framework with prefill--decode disaggregation.
As shown in Figure~\ref{fig:h2d}, prefix-cache fetching accounts for up to 70\% of total TTFT for a 64\,k-token cache hit on Qwen-7B-Chat~\cite{bai2023qwen}.
This fraction increases with context length and depends on model architecture, including hidden size, number of layers, attention heads, and KV heads.
Moreover, in real-world applications such as multi-turn conversations, code generation, and retrieval-augmented generation, context lengths often exceed 100\,k tokens~\cite{jin2024ragcache}, further exacerbating the impact of prefix-cache transfer latency.

\textbf{Model wake-up and switching latency is dominated by weight transfer.}
In vLLM's Sleep Mode (Level 1), a model instance can release its weights from GPU memory and later reload them, enabling memory reuse across serving workloads.
The sleep path copies weights from GPU memory to host memory via D2H transfers, and the wake-up path copies them back via H2D transfers.
We evaluate this transfer overhead across models~\cite{bai2023qwen,yang2025qwen3} of different sizes.
As shown in Figure~\ref{fig:sleep03}, data transfer increasingly dominates total sleep/wake latency as model size grows---from roughly 40--50\% at 0.6B parameters to over 95\% at 32B---and the largest model takes $\sim$2.5\,s to wake or switch, well above sub-second TTFT targets.

\subsection{Latent aggregate bandwidth in multi-GPU servers} \label{Need-multipath}
As outlined in \S\ref{sec:intro}, peer GPUs provide additional physical paths between host memory and a target GPU: a peer can read data over its own PCIe link and forward it to the target through NVLink.
On an 8-GPU node, the upper-bound host-read capacity can therefore grow from one 64\,GB/s PCIe link toward the aggregate capacity of multiple PCIe links, subject to host DRAM and inter-socket limits (Table~\ref{tab:performance}).

The availability of these relay paths is a runtime property.
During cold-start loading, model wake-up, and KV cache prefetching, a large transfer may dominate the critical path while some peer PCIe links, NVLink ports, or copy engines have spare capacity.
During compute-bound decoding or asymmetric prefill/decode execution, spare relay capacity may also appear intermittently.
Conversely, when all GPUs are simultaneously busy with communication or transfers, little relay capacity is available.
MMA is designed for this opportunistic setting: it expands the path set available to each host--GPU transfer and uses only paths that can make progress.
When a relay path is busy, MMA's path selector routes around it or falls back to the direct PCIe path.

\subsection{Multipath memory access challenges} \label{sec:Opportunity and Challenges}

The bandwidth headroom identified above makes multipath transfer attractive, but CUDA's execution model imposes constraints that make exploiting it non-trivial.
A CUDA program expresses work as \emph{tasks} (kernels, memory copies, and events) pushed onto one or more \emph{streams}.
Within a stream, tasks execute in strict FIFO order; across streams, partial ordering is established through \emph{events} that one stream records and another waits on.
The GPU scheduler advances by resolving these dependencies: it does not take real-time directives from the CPU.
Two consequences follow.
First, once a task (including \texttt{cudaMemcpyAsync}) is enqueued, its path, timing, and completion semantics are bound to the stream; the CPU cannot revoke, reroute, or re-sequence it after the fact.
Second, CUDA dependencies can only order work represented in the stream graph; completion of work executed outside that graph is not visible to downstream tasks.

These properties, together with the lack of congestion feedback on PCIe, give rise to three specific challenges for multipath transfer:

\noindent\textbf{C1: Static path binding at enqueue time.}
The transfer path is fixed at enqueue time.
The only way to dispatch transfers across multiple paths dynamically is to synchronize with the GPU before each transfer, waiting until the stream reaches the current task so the CPU can decide which path to use.
These round-trips cause CPU--GPU pipeline stalls and defeat the purpose of asynchronous execution.
The challenge is to delay physical path selection until the transfer is ready to execute, without forcing the application or CPU to synchronize before every copy.
Our key insight is that a thin interposition layer can intercept transfers \emph{before} they reach the GPU pipeline and defer path selection to MMA, without modifying hardware or user code~(\S\ref{Sec:Transfer Task Interceptor}).

\noindent\textbf{C2: Representing distributed transfer completion in CUDA streams.}
CUDA streams preserve correctness by treating a memory copy as a single ordered task: downstream kernels and events are released only after that task completes.
This abstraction matches native single-path DMA, where CUDA owns both the copy operation and its completion.

Multipath transfer breaks this one-task/one-path assumption.
A logical host--GPU copy may be divided across the direct PCIe path and several relay paths, each with independent DMA and GPU-to-GPU forwarding.
The logical transfer is complete only when all fragments have arrived at the target, but CUDA streams provide no native primitive for exposing this distributed completion as the completion of the original copy task.

If the stream-visible task completes before all fragments arrive, downstream kernels may read stale data.
If correctness is restored with device-wide synchronization or CPU-side blocking, the system loses CUDA's asynchronous execution benefits.
The challenge is to preserve the original stream dependency semantics while aggregating completion across multiple physical transfer paths.
We observe that a lightweight GPU-side task can serve as a bidirectional synchronization carrier: it blocks the stream until the CPU-side engine confirms completion, then releases downstream work~(\S\ref{Sec:Sync Engine}).

\noindent\textbf{C3: Heterogeneous paths without congestion visibility.}
Unlike data-center networks where ACKs carry ECN bits and RTT measurements, PCIe exposes no explicit congestion signals to software.
Moreover, the available relay paths are heterogeneous: depending on NUMA placement, a path may traverse local or remote host-memory links before reaching a relay GPU, and each relay path also includes a GPU-interconnect hop to the target. These paths differ in latency, bandwidth, and contention behavior.
Static or random traffic splitting can overload a slow path while leaving a fast one underutilized, causing long-tail completion times.
We find that per-destination buffer occupancy serves as an effective implicit congestion signal, enabling load-aware routing without explicit feedback from the interconnect layer~(\S\ref{sec:multipath Transfer Engine}).

\section{MMA Design}

\subsection{Overview}
\begin{figure}[t]
\centering
\includegraphics[width=\columnwidth]{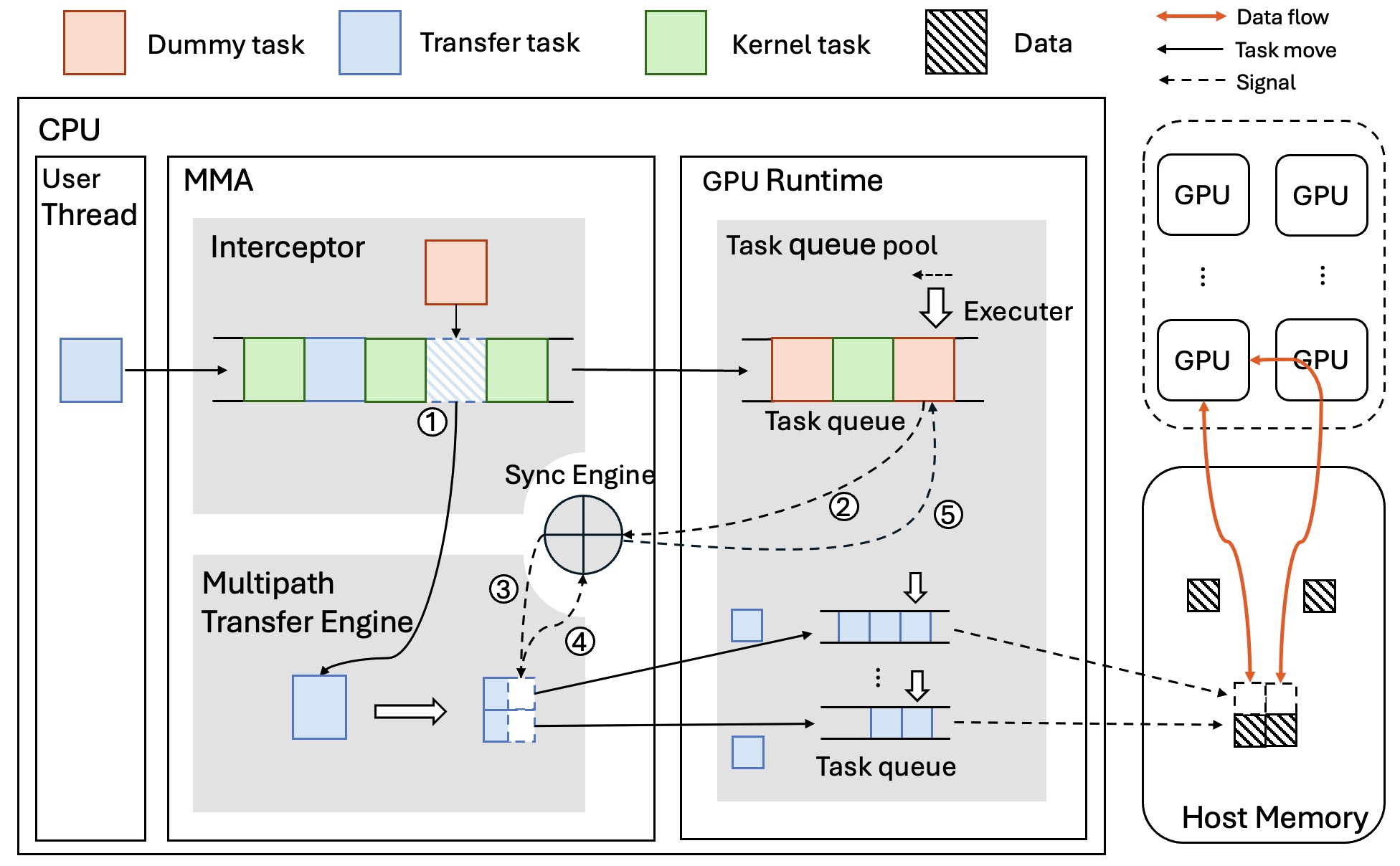}
\vspace{-8pt}
\caption{Overview of MMA.}
\vspace{-5pt}
\label{fig:overview}
\end{figure}

MMA interposes on CUDA's transfer API to reroute host--GPU data movement through multiple PCIe paths, using peer GPUs as relay nodes when relay capacity is available. Its design goal is to maximize aggregate transfer bandwidth while preserving the semantics of existing CUDA applications and retaining the native direct path as the correctness and performance fallback. As shown in Figure~\ref{fig:overview}, MMA comprises three components:

\begin{itemize}[leftmargin=*,nosep]
\item \textbf{Transfer Task Interceptor} (\S\ref{Sec:Transfer Task Interceptor}) records each intercepted host-device copy as a Transfer Task and, for asynchronous copies, replaces the stream-visible copy with a lightweight \emph{Dummy Task}.
\item \textbf{Sync Engine} (\S\ref{Sec:Sync Engine}) keeps the Dummy Task's lifecycle synchronized with the real transfer for asynchronous copies, so that downstream GPU tasks see correct completion ordering.
\item \textbf{Multipath Transfer Engine} (\S\ref{sec:multipath Transfer Engine}) splits each transfer into micro-tasks, selects direct and relay paths using implicit queue backpressure, and launches the actual data movement.
\end{itemize}

\noindent\textbf{End-to-end workflow.}
For an asynchronous copy, when a CUDA memory-copy call is intercepted, MMA records the payload as a Transfer Task and replaces the original stream-visible copy with a Dummy Task~(\ding{172}).
When the stream reaches the Dummy Task, the Sync Engine observes that the original copy point is active~(\ding{173}) and enables the Transfer Engine to dispatch the payload over direct and relay paths~(\ding{174}).
After all micro-tasks complete, the Transfer Engine reports completion to the Sync Engine~(\ding{175}), which releases the Dummy Task and lets CUDA's normal stream ordering resume downstream work~(\ding{176}).

\subsection{Transfer Task Interceptor} \label{Sec:Transfer Task Interceptor}
CUDA binds a memory copy to the target GPU's PCIe path when the copy is submitted.
If MMA let the native copy enter the CUDA pipeline, the transfer would already be committed to the single direct path before the runtime could recruit relay paths.
Rather than requiring hardware changes, MMA interposes at the CUDA memory-copy boundary and hooks both asynchronous and synchronous host-device copy calls.
For asynchronous copies, the Interceptor replaces the stream-visible copy with a Dummy Task, a lightweight placeholder dispatched to the GPU's task queue.
For synchronous copies, MMA reuses the same Transfer Task and Transfer Engine machinery but blocks the calling thread until the real transfer completes, preserving the blocking semantics of the original call.
In both cases, the real payload is recorded as a Transfer Task and dispatched only when it is safe to choose direct or relay paths according to runtime path availability.

Because multipath relay incurs non-trivial setup overhead, MMA includes a \emph{fallback mechanism}: transfers smaller than a configurable threshold bypass the multipath engine and fall back to native single-path copies. We empirically determine the optimal threshold in \S\ref{Sec: Deep Dive}.
This threshold also naturally filters out small control messages.
Furthermore, MMA only applies this interposition to copies between host and device address spaces.
It does not intercept GPU\nobreakdash-to\nobreakdash-GPU copies or NCCL's internal communication, which use separate code paths (peer-to-peer DMA or kernel-based collectives).

\subsection{Sync Engine} \label{Sec:Sync Engine}

For asynchronous copies, MMA replaces the native stream-visible transfer with a Dummy Task, so every downstream GPU operation that originally depended on the transfer now depends on this placeholder.
The Dummy Task is not a new CUDA primitive; MMA implements it as two stream-ordered operations: a host callback that marks the original copy point as active, and a lightweight spin kernel that keeps the stream blocked until the real multipath transfer completes.
The Sync Engine coordinates these operations to keep the placeholder alive \emph{exactly} as long as the real transfer is in flight: releasing too early exposes stale memory; holding too long stalls the pipeline.

\noindent\textbf{Why simpler alternatives fall short.}
CUDA offers several CPU--GPU notification primitives, but none satisfies MMA's requirement of \emph{bidirectional} synchronization: the stream must (1)~notify the CPU that the Dummy Task has started, \emph{and} (2)~block until the CPU confirms that all micro-tasks have landed:
\begin{itemize}[leftmargin=*,nosep]
\item \texttt{cudaDeviceSynchronize} blocks \emph{all} streams on the device, destroying concurrency between independent transfers and compute.
\item CUDA host callbacks such as \texttt{cudaLaunchHostFunc} provide stream$\to$CPU notification only; they cannot make the stream \emph{wait} for a CPU-side signal.
\item Polling a device-memory flag from the CPU requires repeated \texttt{cudaMemcpy} round-trips, adding tens of microseconds per poll and saturating the PCIe control path.
\end{itemize}

\noindent\textbf{Spin-kernel synchronization.}
MMA combines the two operations above to achieve bidirectional signaling without driver modifications.
When the stream reaches the callback, it notifies the Sync Engine that the original copy point is active, allowing the corresponding multipath transfer to begin~(stream$\to$CPU direction).
The following spin kernel waits on a GPU-visible completion flag, preventing the Dummy Task from completing prematurely~(CPU$\to$stream direction).
Once the Transfer Engine finishes all micro-tasks, the Sync Engine sets the flag; the spin kernel observes the update and exits, releasing downstream GPU work.
Because the spin kernel is intentionally lightweight, its SM footprint is negligible; we verify this claim experimentally in \S\ref{Sec:impl}.

\subsection{Multipath Transfer Engine} \label{sec:multipath Transfer Engine}

As illustrated in Figure~\ref{fig:multipath}, the Multipath Transfer Engine has three components: the \emph{Task Manager} splits each transfer into micro-tasks and manages the micro-task queue; the \emph{Path Selector} moves micro-tasks into per-link outstanding queues using implicit queue backpressure; and the \emph{Task Launcher} dispatches the actual GPU DMA operations.

\subsubsection{Task Manager} \label{Sec:Task Manager}

\begin{figure}[t]
\centering
\includegraphics[width=\columnwidth]{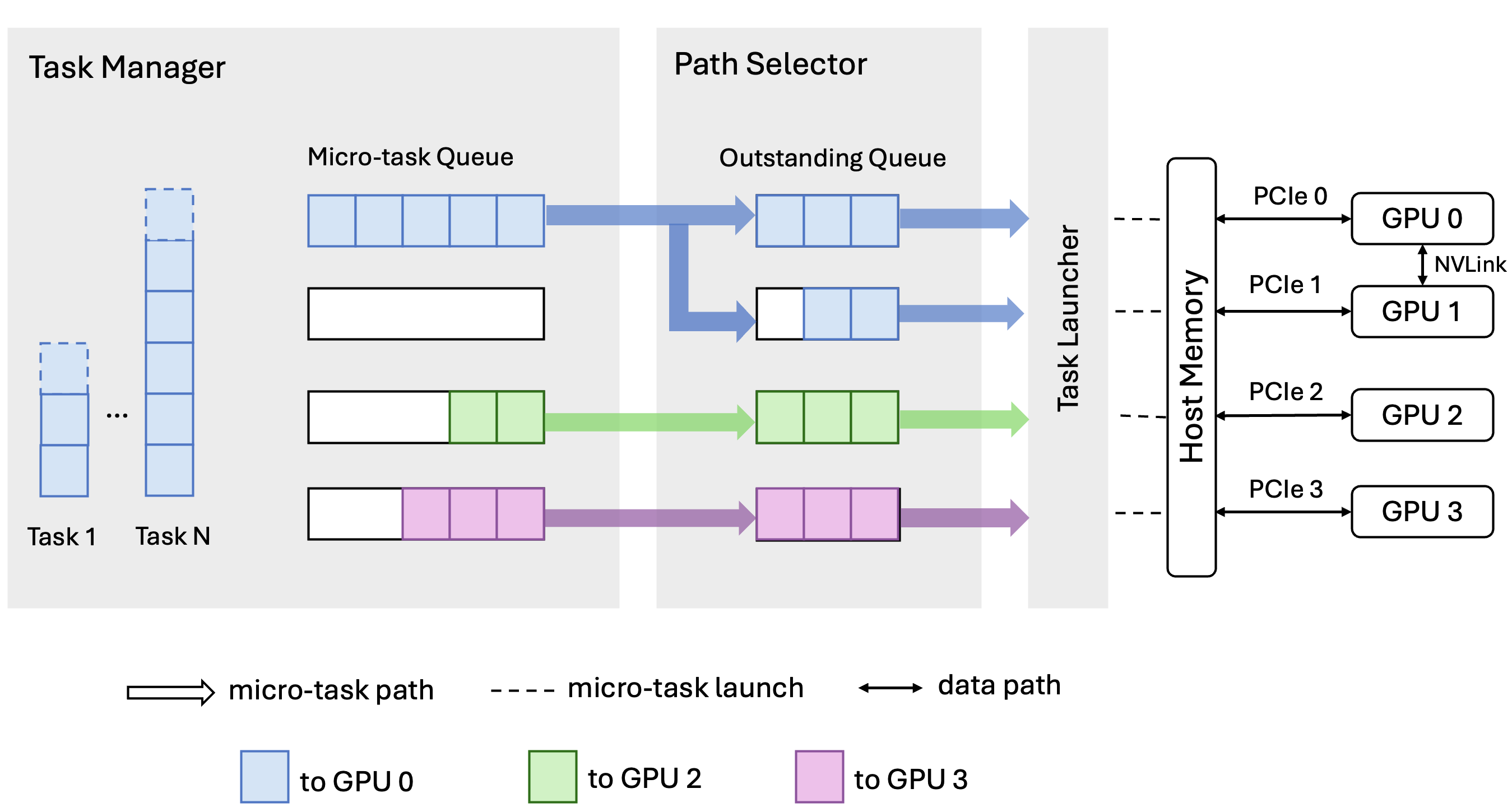}
\vspace{-10pt}
\caption{Multipath Transfer Engine. Colors denote the destination GPU of each micro-task; the Path Selector moves micro-tasks from the micro-task queue into per-link outstanding queues.}
\label{fig:multipath}
\vspace{-5pt}
\end{figure}

Upon receiving the start signal from the Sync Engine, the Task Manager divides the original transfer into fixed-size micro-tasks and places them into the micro-task queue shown in Figure~\ref{fig:multipath}.
Each micro-task is tagged with its destination GPU, which the figure denotes by color.
The chunk size balances two concerns: chunks that are too small fail to saturate link bandwidth, while chunks that are too large reduce load-balancing granularity; we empirically determine the sweet spot in \S\ref{Sec: Deep Dive}.
After all micro-tasks complete, the Task Manager notifies the Sync Engine, which releases the corresponding stream-visible placeholder for asynchronous copies or wakes the blocked caller for synchronous copies.

This destination-tagged micro-task queue lets the Path Selector prioritize direct-path transfers (those whose destination matches the GPU attached to a PCIe link), minimizing unnecessary NVLink relay hops.

\subsubsection{Path Selector}

MMA maintains one \emph{outstanding queue} per PCIe link, statically bound to the corresponding GPU.
The Path Selector fills these outstanding queues by pulling work from the micro-task queue, and the Task Launcher dispatches queued micro-tasks onto the corresponding links.
If a micro-task's destination GPU matches the outstanding queue's GPU, it travels via the direct PCIe path; otherwise it takes a relay path (PCIe to the relay GPU, then NVLink to the target).

Rather than actively probing link utilization, MMA uses outstanding-queue backpressure as an implicit congestion signal: if transfers issued on a path complete slowly, the path's outstanding queue remains full and stops pulling new work; faster paths drain their queues and pull more micro-tasks from the micro-task queue.

\noindent \textbf{Direct path first.}
Relay paths consume NVLink bandwidth in addition to PCIe.
To minimize this overhead, each outstanding queue prioritizes micro-tasks destined for its own GPU before pulling relay work from other queues (Figure~\ref{fig:multipath}); we quantify the benefit in \S\ref{Sec: Deep Dive}.

\noindent \textbf{Relay task scheduling.}
When an outstanding queue has drained direct-path micro-tasks for its own GPU, it steals relay work from the destination with the largest amount of remaining data in the micro-task queue.
This longest-remaining-destination policy maximizes the fraction of data delivered via direct paths across all GPUs, reducing aggregate NVLink consumption.

\noindent \textbf{Contention with background traffic.}
If a relay link is contended by co-running workloads, its outstanding queue backs off: it waits until the queue depth drops below a threshold before pulling new micro-tasks, yielding bandwidth to latency-sensitive background traffic.
When no contention is detected, the queue pulls immediately to maximize throughput.

\subsubsection{Task Launcher}

The Task Launcher dispatches micro-tasks to the GPU via the CUDA runtime.
For a \emph{direct} micro-task, it issues a single H2D (or D2H) DMA transfer.
For an H2D relay micro-task, it issues an H2D copy to the relay GPU followed by a GPU-to-GPU P2P copy to the target.
For a D2H relay micro-task, it performs the reverse order: a P2P copy from the target to the relay GPU followed by a D2H copy to host memory.
In both directions, explicit stream dependencies preserve ordering between the two stages.

\begin{figure}[htbp]
\centering
\includegraphics[width=\columnwidth]{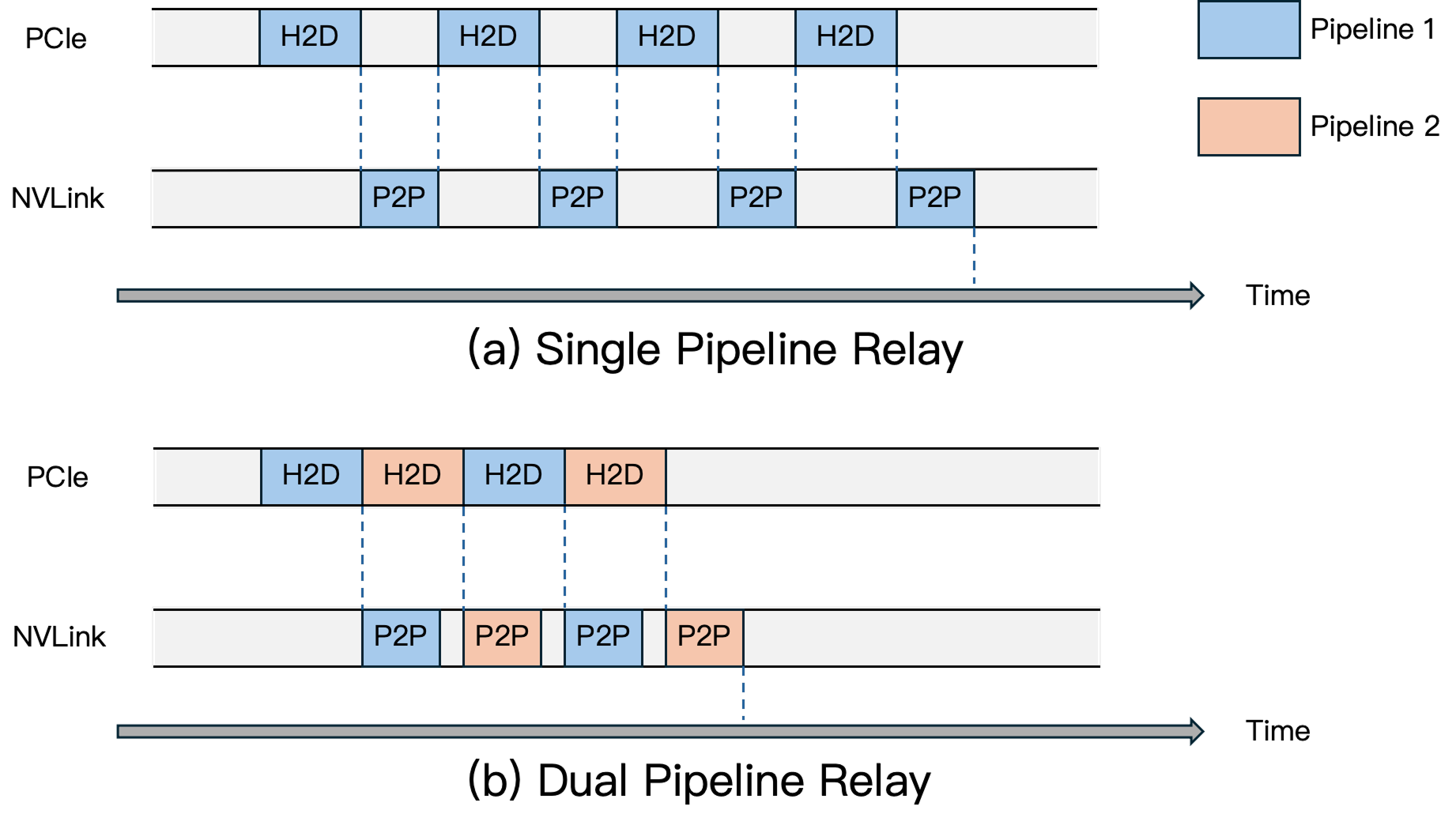}
\vspace{-10pt}
\caption{Dual-pipeline relay overlaps PCIe ingress/egress with NVLink forwarding to improve relay utilization.}
\vspace{-5pt}
\label{fig:dualrelaystream}
\end{figure}

\noindent \textbf{Dual-pipeline relay.}
A na\"ive single-pipeline relay leaves one stage idle while the other stage transfers data between host memory, the relay GPU, and the target GPU (Figure~\ref{fig:dualrelaystream}(a)).
MMA eliminates these bubbles with a \emph{dual-pipeline} design (Figure~\ref{fig:dualrelaystream}(b)): two relay streams per GPU overlap the PCIe stage and the NVLink stage in a ping-pong fashion, keeping both links busy.

\noindent \textbf{GPU overhead.}
H2D, D2H, and P2P transfers use the GPU's DMA copy engines, not its SMs. The only SM footprint is the spin kernel on the target GPU (\S\ref{Sec:Sync Engine}), which occupies a single thread block ($<$1\% of H20's 132 SMs). Relay therefore does not meaningfully compete with compute workloads.
Each relay stream requires a dedicated buffer of one chunk (5\,MB by default, see \S\ref{Sec: Deep Dive}); with two streams and bidirectional traffic the total memory overhead per GPU is 20\,MB---negligible on modern GPUs.

%% ============================================================
%% SECTION 5: Implementation
%% ============================================================
\section{Implementation}\label{Sec:impl}
MMA is implemented atop CUDA 12.8~\cite{cuda} in approximately 3\,000 lines of C++. It supports both transparent and explicit integration: unmodified applications use \texttt{LD\_PRELOAD} interposition on \texttt{cudaMemcpy}/\texttt{cudaMemcpyAsync}, while framework developers can call MMA through Python and C++ APIs.

This section details two key aspects of the implementation: the Multipath Transfer Engine and the Spin Kernel.

\noindent \textbf{Multipath Transfer Engine.} MMA maintains separate H2D and D2H engine instances. In the default flow-control mode, each engine creates per-GPU worker threads for transfer dispatch, completion tracking, and queue monitoring: a \emph{transfer thread}, a \emph{synchronization thread}, and a \emph{monitor thread}. The implementation also supports a centralized dispatch mode in which one transfer worker dispatches work across GPUs while synchronization and monitoring remain per-GPU.

The transfer thread drives one per-link outstanding queue. When the queue has capacity, it invokes the Path Selector to pull a micro-task from the micro-task queue, dispatches the selected micro-task through the Task Launcher, records a CUDA event as a synchronization point, and submits the corresponding synchronization task to the synchronization thread.

The synchronization thread calls \texttt{cudaEventSynchronize} to block until the dispatched micro-task batch completes. It then retires the completed work from the outstanding queue and checks whether all chunks of the original transfer have been delivered.

If all chunks are complete, the synchronization thread notifies the Sync Engine, signaling that the transfer has finished.

\noindent \textbf{Spin Kernel.} 
For asynchronous copies, the spin kernel synchronizes the Dummy Task with the multipath transfer.
The mechanism works as follows:
(1)~A control flag \texttt{h\_flag} is allocated in pinned host memory with the \texttt{cudaHostAllocMapped} attribute and mapped to a device-accessible pointer \texttt{d\_flag} via \texttt{cudaHostGetDevicePointer}.
This creates a mapped host-memory flag that the GPU can read without explicit \texttt{cudaMemcpy} calls.
(2)~When the stream reaches the Dummy Task, the spin kernel enters a loop that reads \texttt{d\_flag} using a cache-global load intrinsic (\texttt{\_\_ldcg}) to avoid stale L1-cached values and repeatedly observe host-side updates.
(3)~Once the Multipath Transfer Engine completes all relay transfers, it sets \texttt{*h\_flag} to~1. The spin kernel observes the update on a subsequent \texttt{\_\_ldcg} load and exits, releasing downstream GPU work.
Between consecutive polls the kernel executes \texttt{\_\_nanosleep(100)}, reducing memory-bus pressure to negligible levels.
In practice, the end-to-end polling latency (CPU store to GPU observation) is on the order of a PCIe round trip ($\sim$1--2\,\textmu s), which is negligible relative to the millisecond-scale multipath transfers.
Because the spin kernel occupies only one thread block (one warp), its resource footprint is small; the main liveness requirement is that the CUDA context remains scheduled, which is common to GPU-side synchronization primitives.

\noindent \textbf{Deployment and Portability.}
MMA's transparent mode targets \texttt{cudaMemcpy}/\texttt{cudaMemcpyAsync}, the common denominator of host--GPU data movement in frameworks such as vLLM~\cite{kwon2023efficient} and LMCache~\cite{lmcache}; no source-level changes are needed because these frameworks commonly implement host--GPU KV cache and weight movement through CUDA memory-copy APIs or wrappers around them.
The explicit integration path uses the same C++ runtime: the Python API is packaged as an importable module backed by pybind11, enabling Python-based serving frameworks to invoke MMA without process-wide interposition.
At startup, MMA queries the GPU topology via NVML and automatically identifies relay candidates based on NUMA affinity and NVLink/xGMI connectivity, eliminating manual configuration.
All runtime parameters (relay GPU list, chunk size, bandwidth threshold, and flow-control mode) are exposed as environment variables.
The core mechanism---CUDA peer-to-peer DMA, \texttt{cudaHostAllocMapped} zero-copy, and \texttt{\_\_ldcg} polling---requires only standard CUDA 11+ APIs and PCIe/NVLink P2P support, making MMA portable in principle to PCIe/NVLink GPU servers such as A100, H100, and H200, subject to P2P support and topology constraints.

%% ============================================================
%% SECTION 6: Evaluation
%% ============================================================
\section{Evaluation}

We first use targeted microbenchmarks to characterize MMA's transfer performance and design choices, then evaluate overall performance on end-to-end LLM serving workloads, and finally present deep-dive analyses of key design components.

\textbf{Evaluation testbed.}
All experiments run on a dual-socket AMD EPYC 9654 server with eight NVIDIA H20 GPUs interconnected via NVLink 4.0 (18 links per GPU) and NVSwitch; each GPU is attached to the host through PCIe 5.0 $\times$16.
The host provides 24-channel DDR5-4800 per socket and the two sockets are connected via 4$\times$xGMI3.
Software: CUDA 12.8, Ubuntu 22.04, vLLM 0.8.0~\cite{kwon2023efficient}, LMCache~\cite{lmcache}.

\textbf{Baseline.} Native CUDA memory copy with static binding to a single PCIe link for all CPU--GPU communication.

\subsection{Microbenchmarks}
\label{sec:microbenchmarks}

\subsubsection{Bandwidth improvement}
\label{sec:micro-bandwidth improvement}

We evaluate MMA's bandwidth improvement over the baseline for host-to-GPU transfers across various message sizes.

\textbf{Setup.}
We sweep the message size from 1\,KB to 8\,GB for both H2D and D2H transfers using pinned host buffers. For each point, we run warm-up iterations, time the transfer with CUDA events and stream synchronization, and report effective bandwidth as message size divided by completion time averaged over repeated runs. The native baseline uses the same buffer allocation and timing method with \texttt{cudaMemcpyAsync} on the target GPU's direct PCIe path.

\begin{figure}[ht]
    \centering

    \subfloat[H2D transfer bandwidth]{
        \includegraphics[width=0.48\columnwidth]{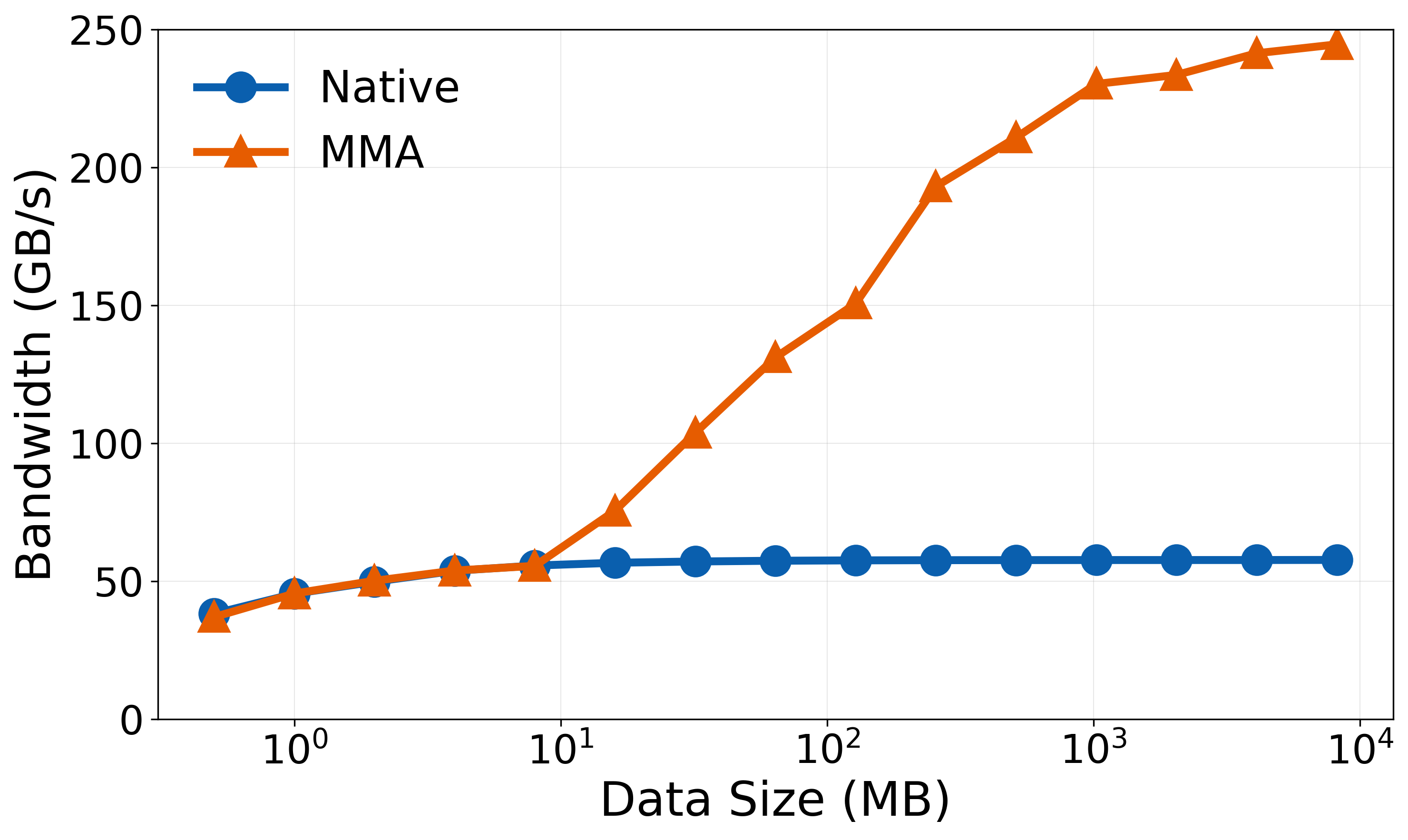}%
        \label{subfig:h2d}%
    }
    \subfloat[D2H transfer bandwidth]{%
        \includegraphics[width=0.48\columnwidth]{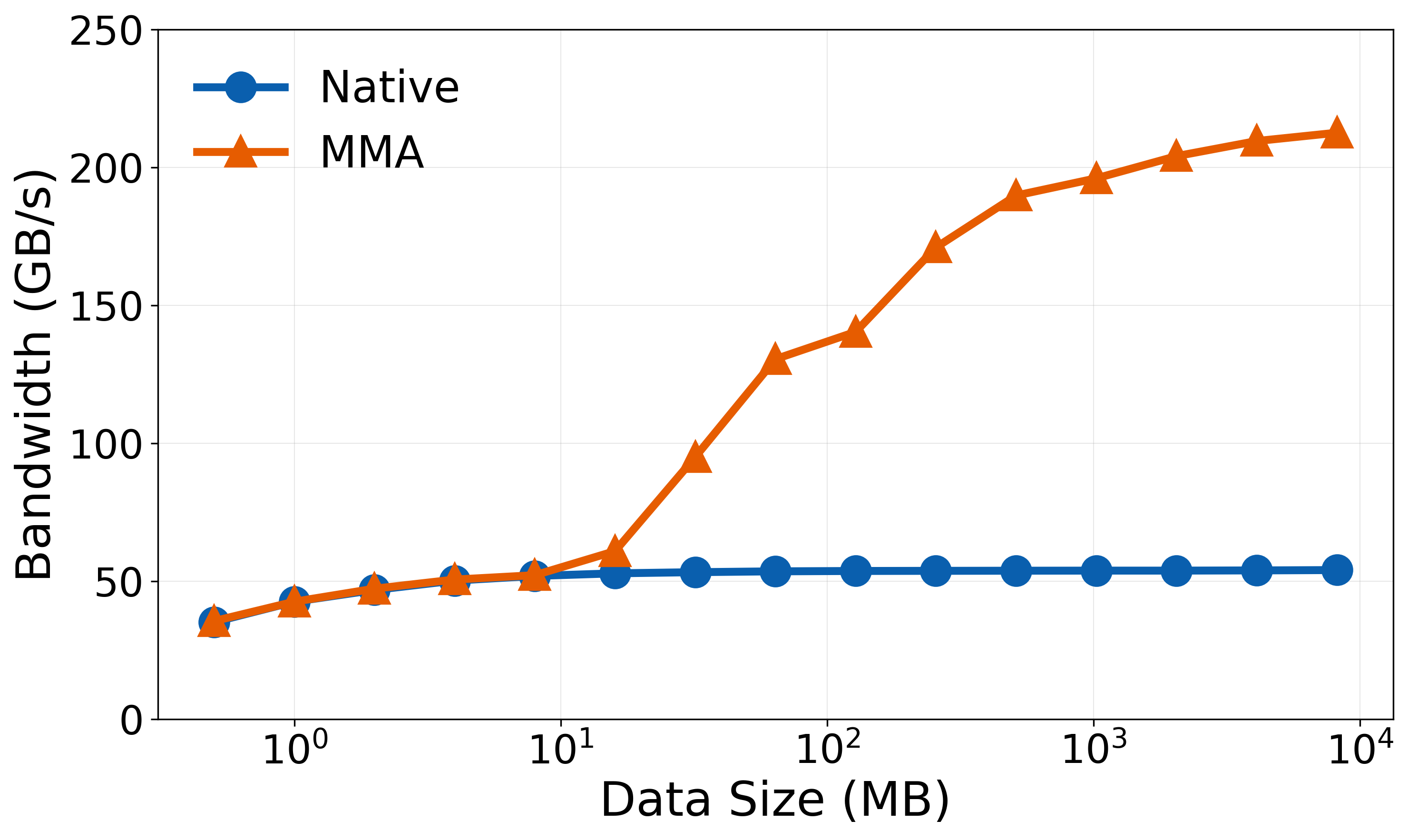}%
        \label{subfig:d2h}%
    }
\vspace{-8pt}
\caption{Bandwidth performance with transfer task size for H2D and D2H transfers.}
\label{fig:bandwidth-dataSize}
\vspace{-5pt}
\end{figure}

As shown in Figure~\ref{fig:bandwidth-dataSize}, MMA begins to outperform the baseline at approximately 10\,MB and approaches its peak bandwidth of 245\,GB/s around 1\,GB.
The baseline saturates at approximately 53\,GB/s, yielding a 4.62$\times$ speedup.
In LLM workloads, intra-node H2D and D2H transfers (e.g., KV caches and model weights) typically range from hundreds of megabytes to tens of gigabytes, well within MMA's effective operating region.

\begin{figure}[ht]
    \centering

    \subfloat[H2D transfer bandwidth]{%
        \includegraphics[width=0.48\columnwidth]{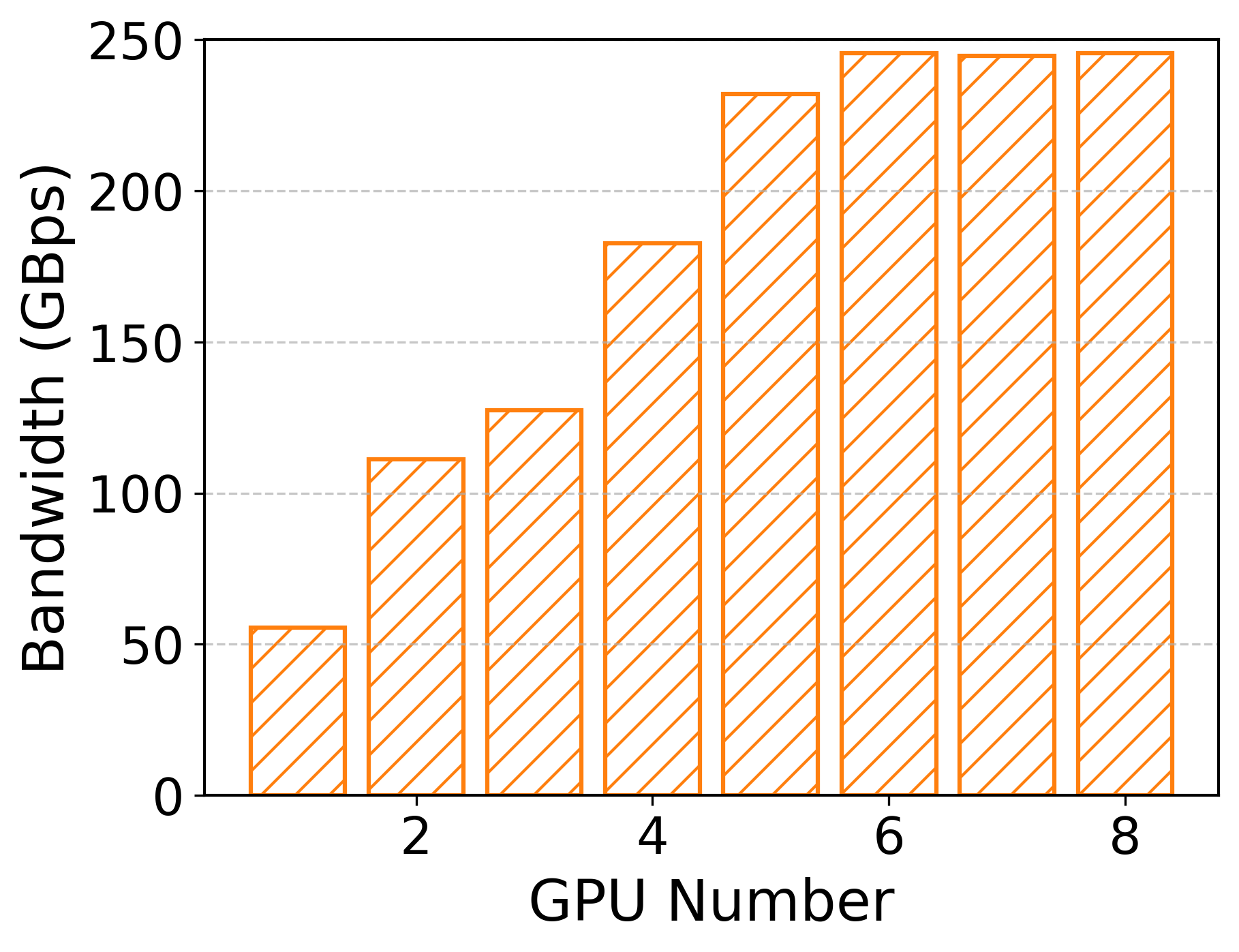}%
        \label{subfig:h2d_path}%
            }
   \subfloat[D2H transfer bandwidth]{%
        \includegraphics[width=0.48\columnwidth]{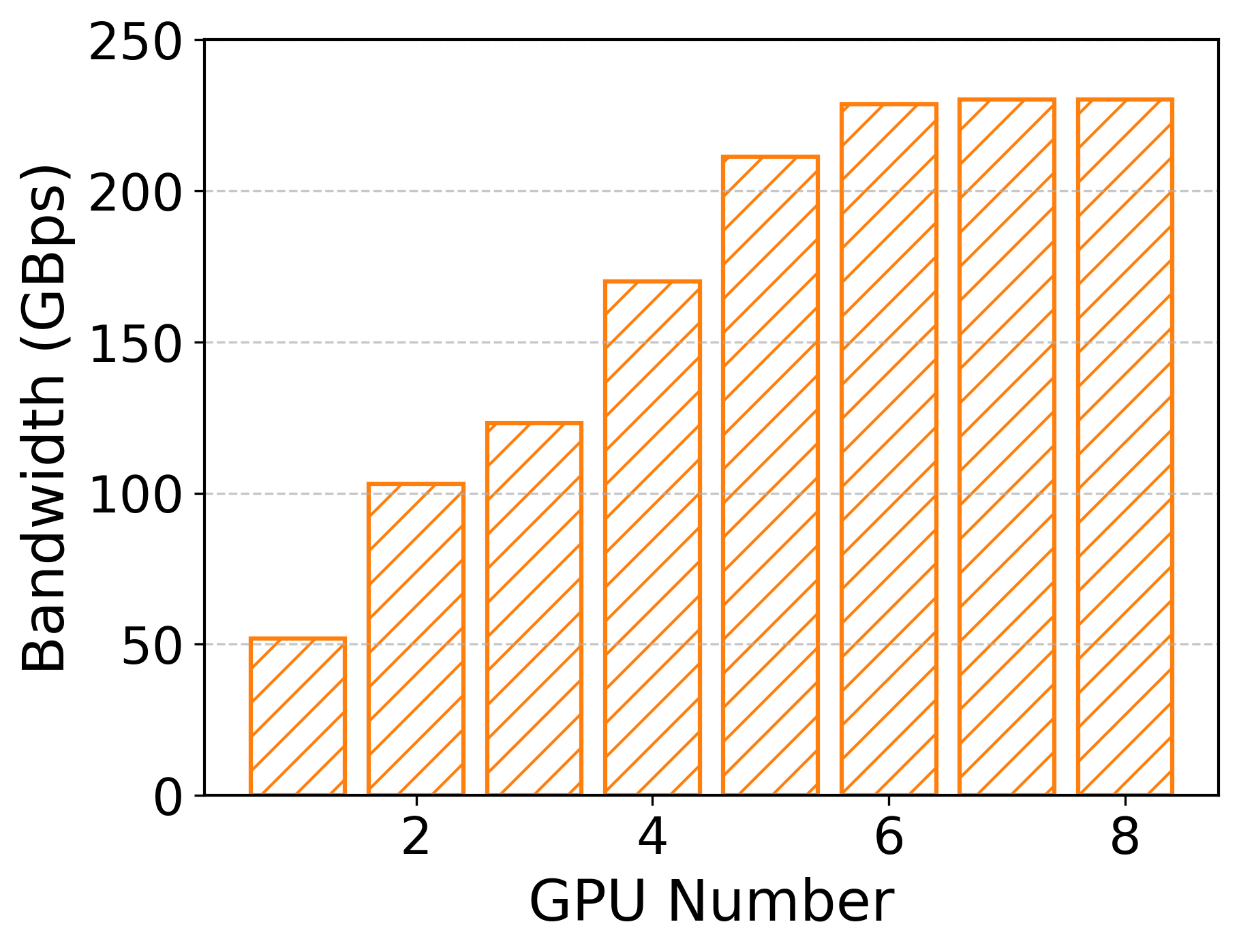}%
        \label{subfig:d2h_path}%
        }
\vspace{-8pt}
\caption{MMA transfer bandwidth versus number of paths.}
\label{fig:bandwidth-path}
\vspace{-8pt}
\end{figure}

Furthermore, Figure~\ref{fig:bandwidth-path} quantifies MMA's sensitivity to relay availability by varying the number of participating peer GPUs.
Bandwidth increases as more relay paths become available and saturates when six GPUs participate in relay.
This is because four GPUs typically reside within a single NUMA node, and cross-NUMA transfers must traverse the inter-socket xGMI3 links (Infinity Fabric), whose aggregate raw one-way bandwidth is approximately 256\,GB/s before protocol overheads.
The measured saturation point of ${\sim}$245\,GB/s suggests that cross-socket bandwidth becomes the residual bottleneck once cross-NUMA relay paths are added.
The resulting 4.62$\times$ speedup (245\,GB/s\,/\,53\,GB/s) falls below the theoretical 8$\times$ because three factors cap throughput before all eight PCIe links can be fully utilized.
First, xGMI3 cross-socket bandwidth limits how much data can reach the remote NUMA node.
Second, relay scheduling overhead (micro-task dispatch, dual-pipeline synchronization) consumes a small fraction of each chunk's transfer time.
Third, DMA copy-engine contention on the target GPU serializes the final NVLink-to-HBM writes from multiple relay streams.

Notably, D2H bandwidth with MMA is consistently lower than H2D.
In the H2D (relay) direction, the first hop (host$\to$relay GPU) and the second hop (relay GPU$\to$target GPU via NVLink) use \emph{different} physical links and can overlap.
In the D2H (relay) direction, the target GPU first sends data to a relay GPU via NVLink, but the relay GPU must then forward the data to the host through \emph{its own} PCIe link.
When multiple relay GPUs share a NUMA node, their outbound PCIe bandwidth to the host aggregates well; however, each relay's PCIe egress is still capped at 64\,GB/s, and the relay GPU must serialize the NVLink-ingress and PCIe-egress stages within its internal DMA engine, reducing effective overlap.

\begin{figure}[htbp]
    \centering

    \subfloat[MMA congestion with Native CUDA]{%
        \includegraphics[width=0.48\columnwidth]{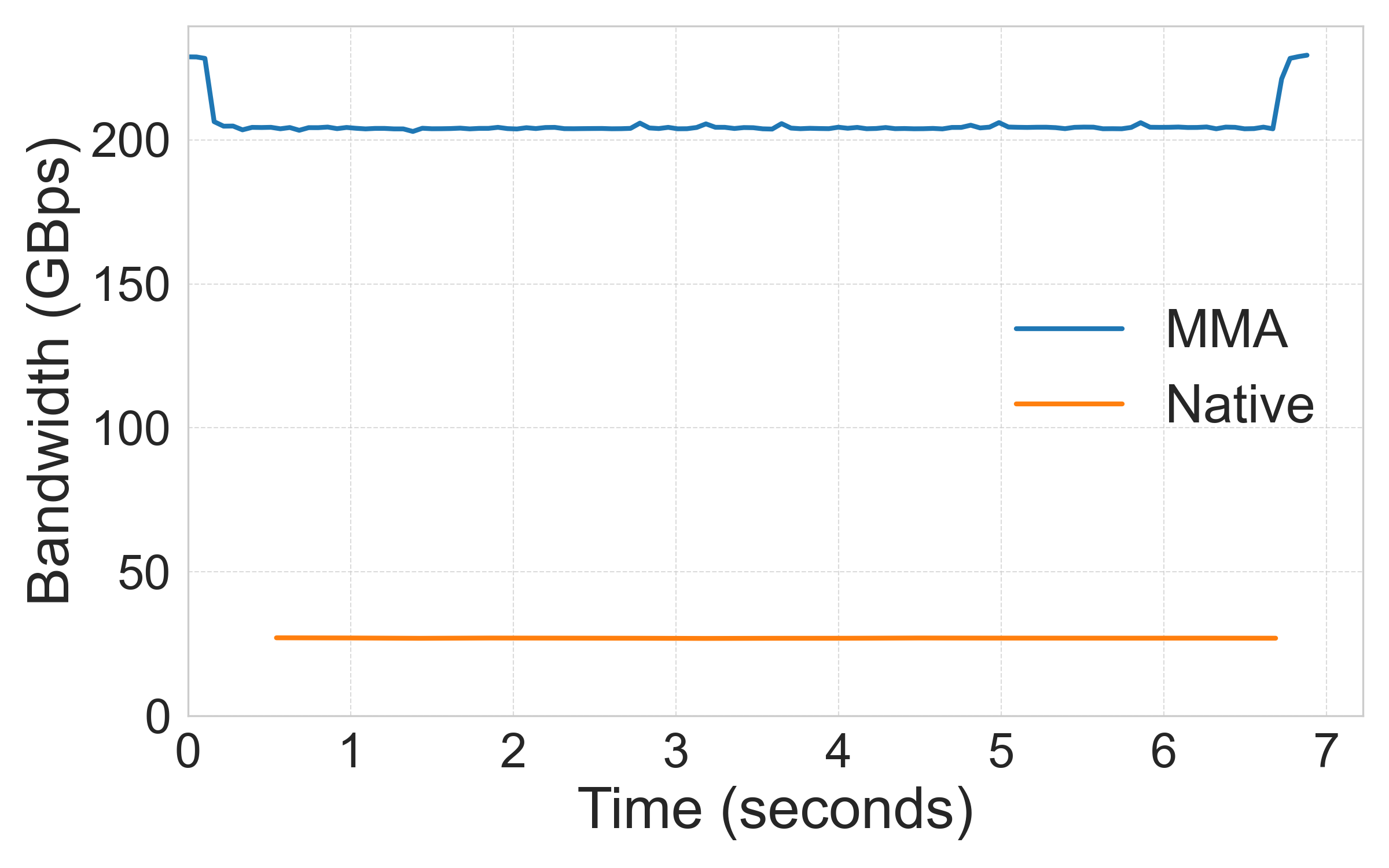}%
        \label{subfig:mma_native}%
    }
    \hfill
    \subfloat[MMA congestion with MMA]{
        \includegraphics[width=0.48\columnwidth]{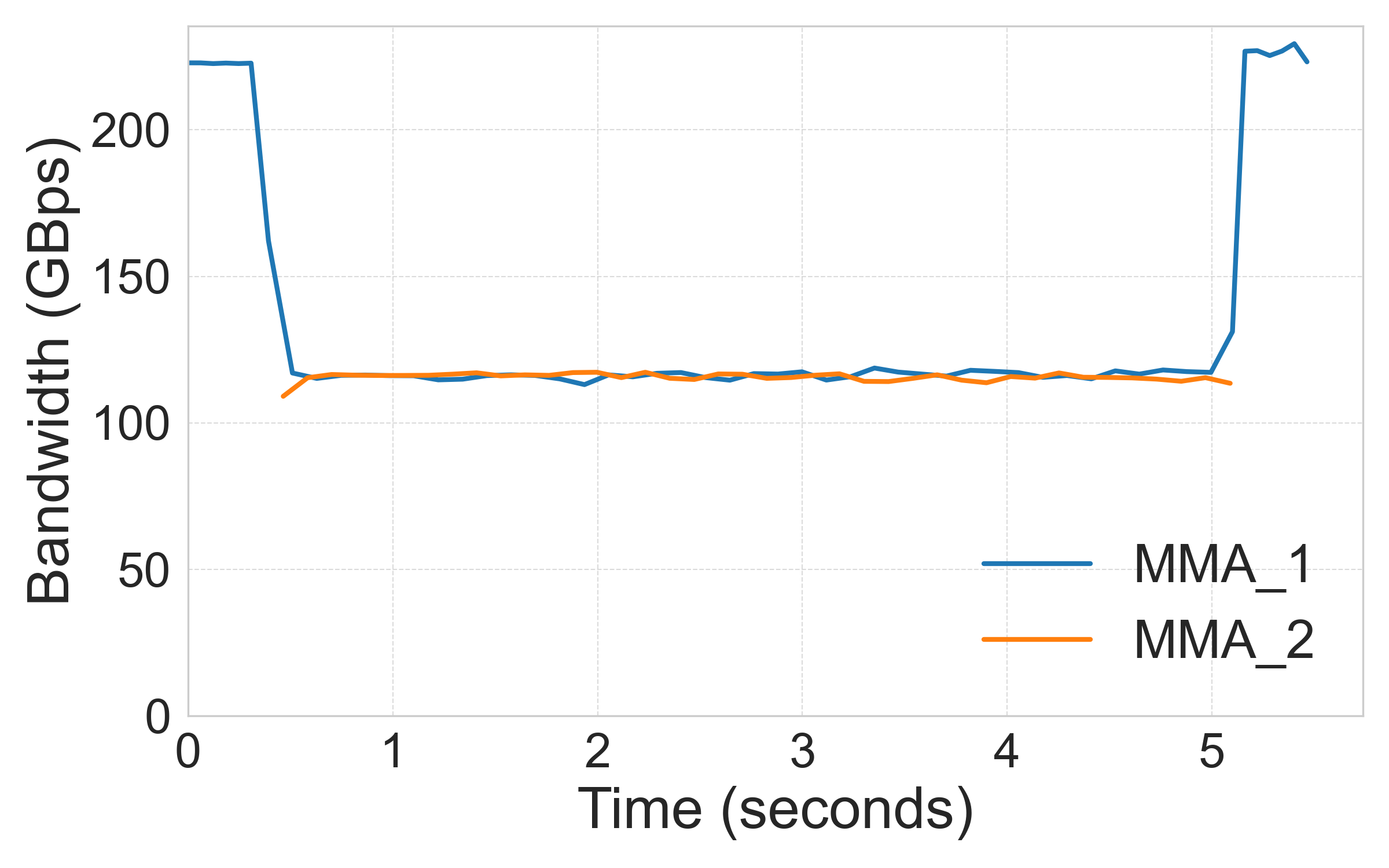}%
        \label{subfig:mma_mma}%
    }
\vspace{-8pt}
\caption{Bandwidth variations during congestion events under MMA and Native transfer methods.}
\label{fig:congestion}
\vspace{-5pt}
\end{figure}

\subsubsection{Robustness and Load Balancing}
\label{sec:design-eval}

We evaluate how robust MMA is under contention and how well it balances load across paths.
Our \emph{baseline} configuration is that both streams use native cudaMemcpyAsync with static binding to fixed PCIe links as shown in Figure~\ref{fig:topology}.

\noindent \textbf{MMA coexistence with native CUDA traffic.}
\label{sec:design-native}We first test how MMA behaves when sharing PCIe bandwidth with native \texttt{cudaMemcpyAsync} traffic.
Here, native copies represent background traffic produced by third-party components (e.g., NIC DMA or applications not using MMA) that pin one direct PCIe link for an extended period.
Figure~\ref{subfig:mma_native} shows the bandwidth allocation when native single-stream CUDA traffic coexists with MMA.
On congested links, MMA reduces the amount of work assigned to the contended path, while non-contended paths continue to contribute bandwidth.
This behavior arises because outstanding-queue backpressure slows pulls from busy paths, and PCIe's internal flow control arbitrates bandwidth between co-running traffic sources.

\noindent \textbf{Coexistence of concurrent MMA flows.}\label{sec:design-mp-vs-mp}
Next, we examine interference between two concurrent MMA flows.
As shown in Figure~\ref{subfig:mma_mma}, the two flows share the available relay capacity without either flow collapsing to the native single-path baseline. Each MMA process maintains its own multipath queue and pull-based scheduler, so although the two flows may contend on shared links, both still achieve bandwidth far exceeding the native baseline.

\noindent \textbf{MMA adapts to dynamic traffic better than static splitting.}
\label{sec:design-lb} To evaluate MMA's adaptability to dynamic intra-node traffic, we compare MMA's pull-based scheduling against static splitting and native single-path transfer, with and without background traffic. For simplicity, we restrict the number of relay paths to two and test two static split ratios: 1:1 and 1:2.

\begin{figure}[htbp]
    \centering
    \begin{minipage}{0.48\columnwidth}
        \centering
        \captionsetup{skip=5pt}
        \includegraphics[width=\linewidth]{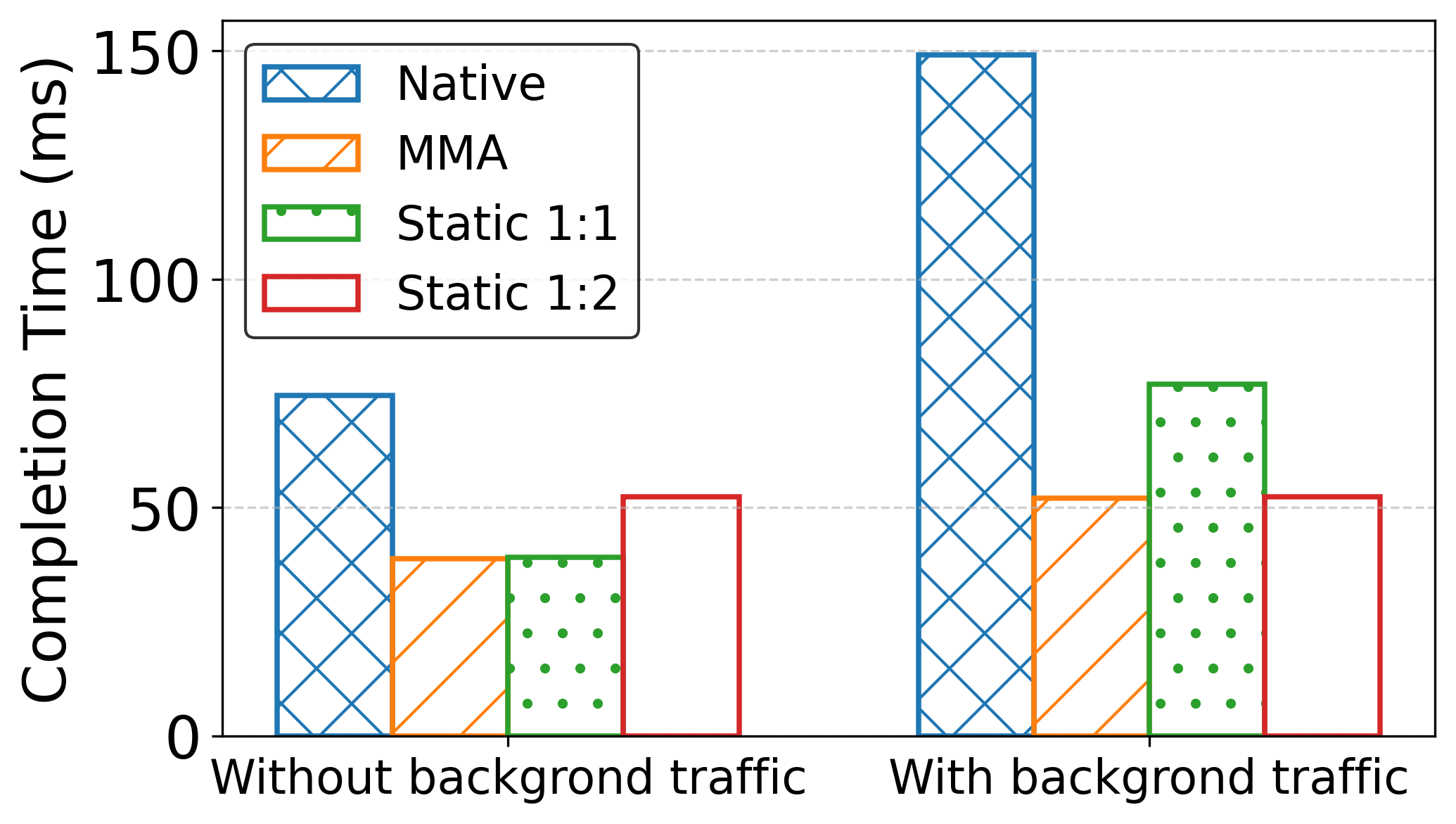}
        \captionof{figure}{Comparison of completion time with and without background traffic.}
        \label{fig:static_mma}
    \end{minipage}
    \hfill
    \raisebox{5pt}{
        \begin{minipage}{0.48\columnwidth}
            \centering
            \captionsetup{skip=0pt}
            \includegraphics[width=\linewidth]{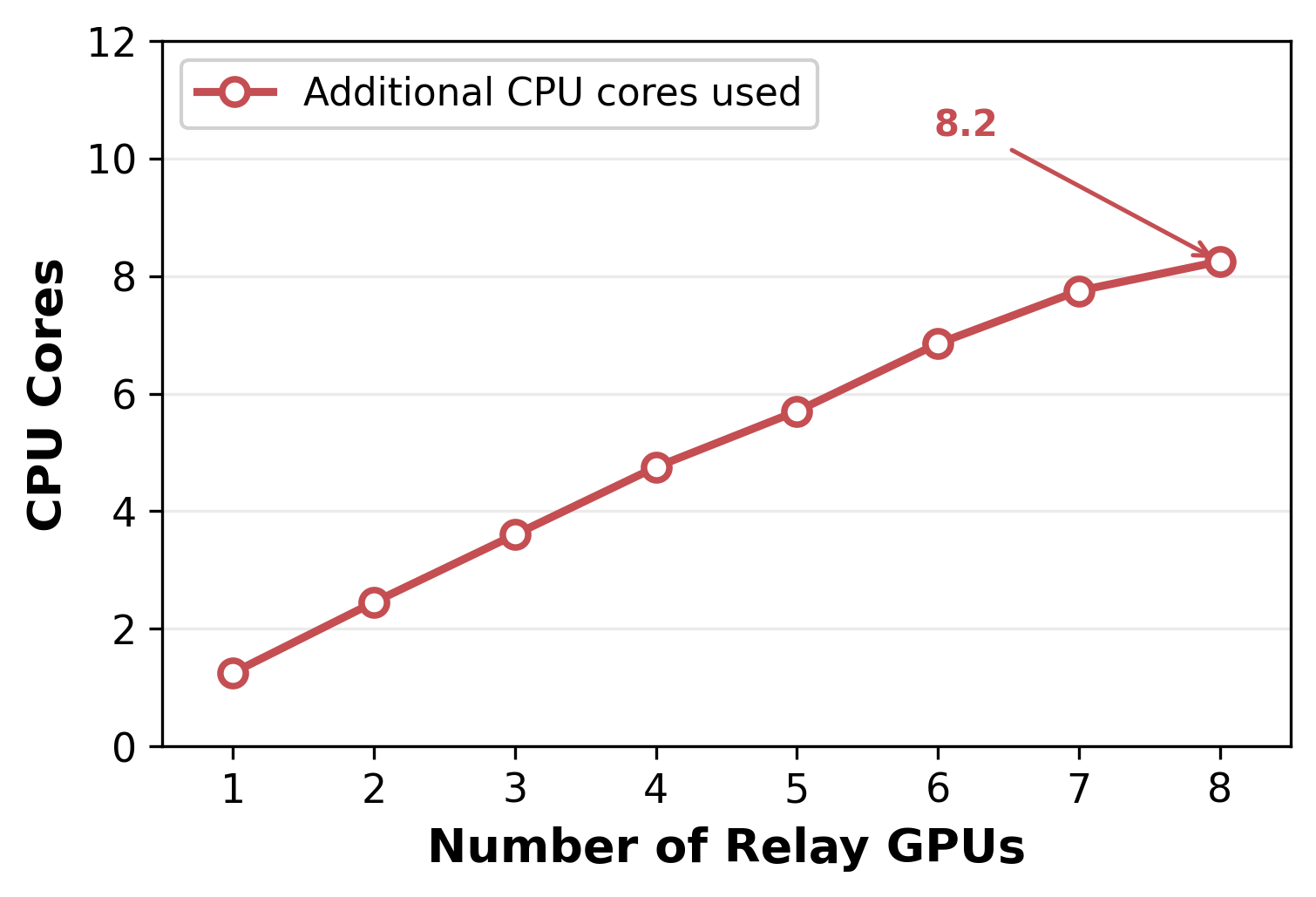}
            \captionof{figure}{Additional CPU cores consumed by MMA (measured via process CPU time).}
            \label{fig:cpu_load_comparison}
        \end{minipage}
    }
    \vspace{-10pt}
\end{figure}

As shown in Figure~\ref{fig:static_mma}, MMA tracks the better static split across both scenarios, while any fixed split performs well only under the traffic pattern it was tuned for.
Without background traffic, the two paths have similar bandwidths, so a 1:1 split is effective.
With background traffic, PCIe flow control reduces the congested path's bandwidth to roughly half, making a 1:2 split the better static choice.
MMA adapts to both conditions automatically, avoiding the need to choose a split ratio in advance.

\begin{figure*}[!t]
\centering
\subfloat[Qwen3-0.6B]{
    \includegraphics[width=0.24\textwidth]{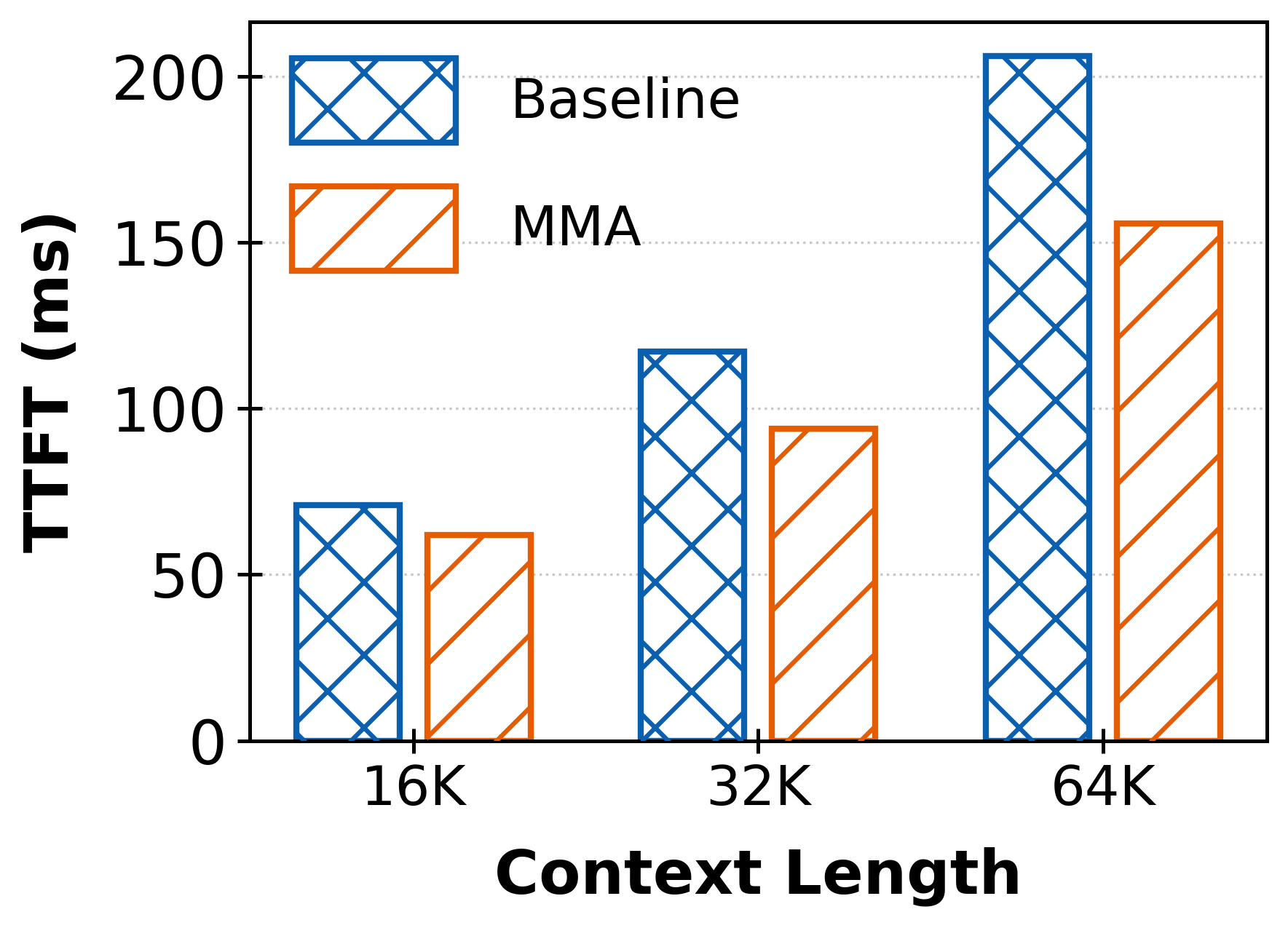}
}
\subfloat[Qwen3-4B]{%
    \includegraphics[width=0.24\textwidth]{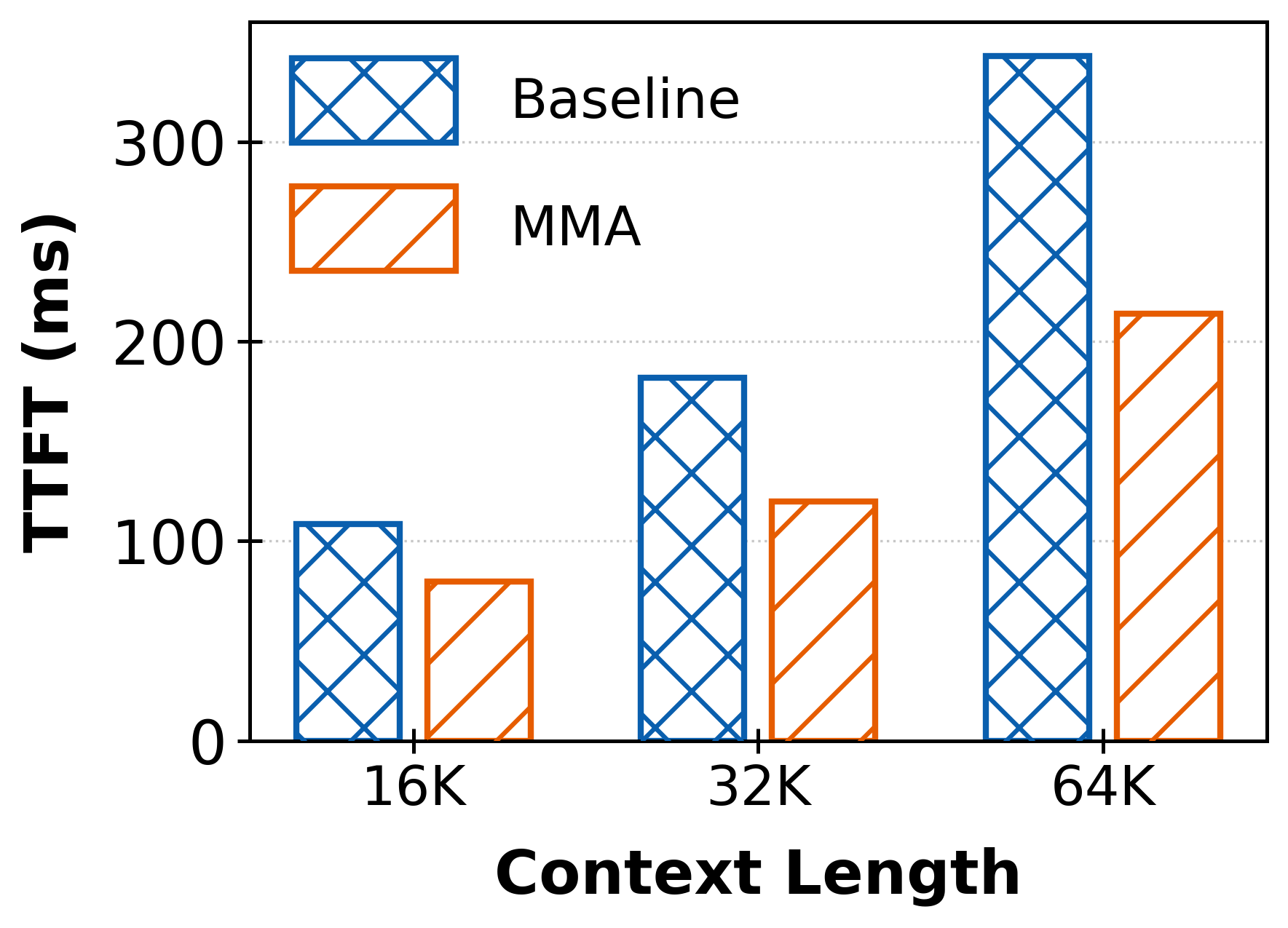}%
    \label{subfig:ttft_b}
}
\subfloat[Qwen-7B-Chat]{%
    \includegraphics[width=0.24\textwidth]{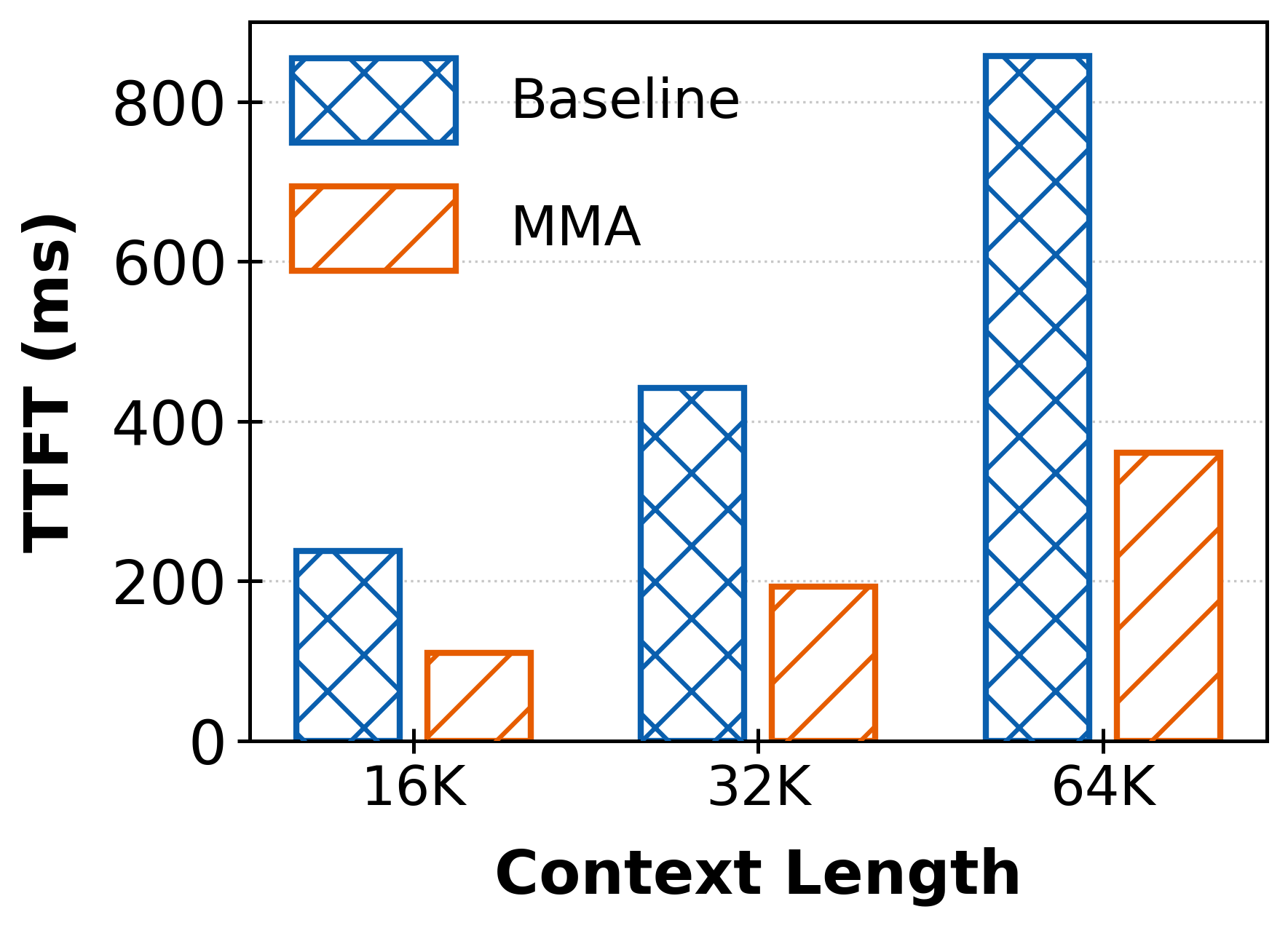}%
    \label{subfig:ttft_c}
}
\subfloat[Qwen3-32B]{%
    \includegraphics[width=0.24\textwidth]{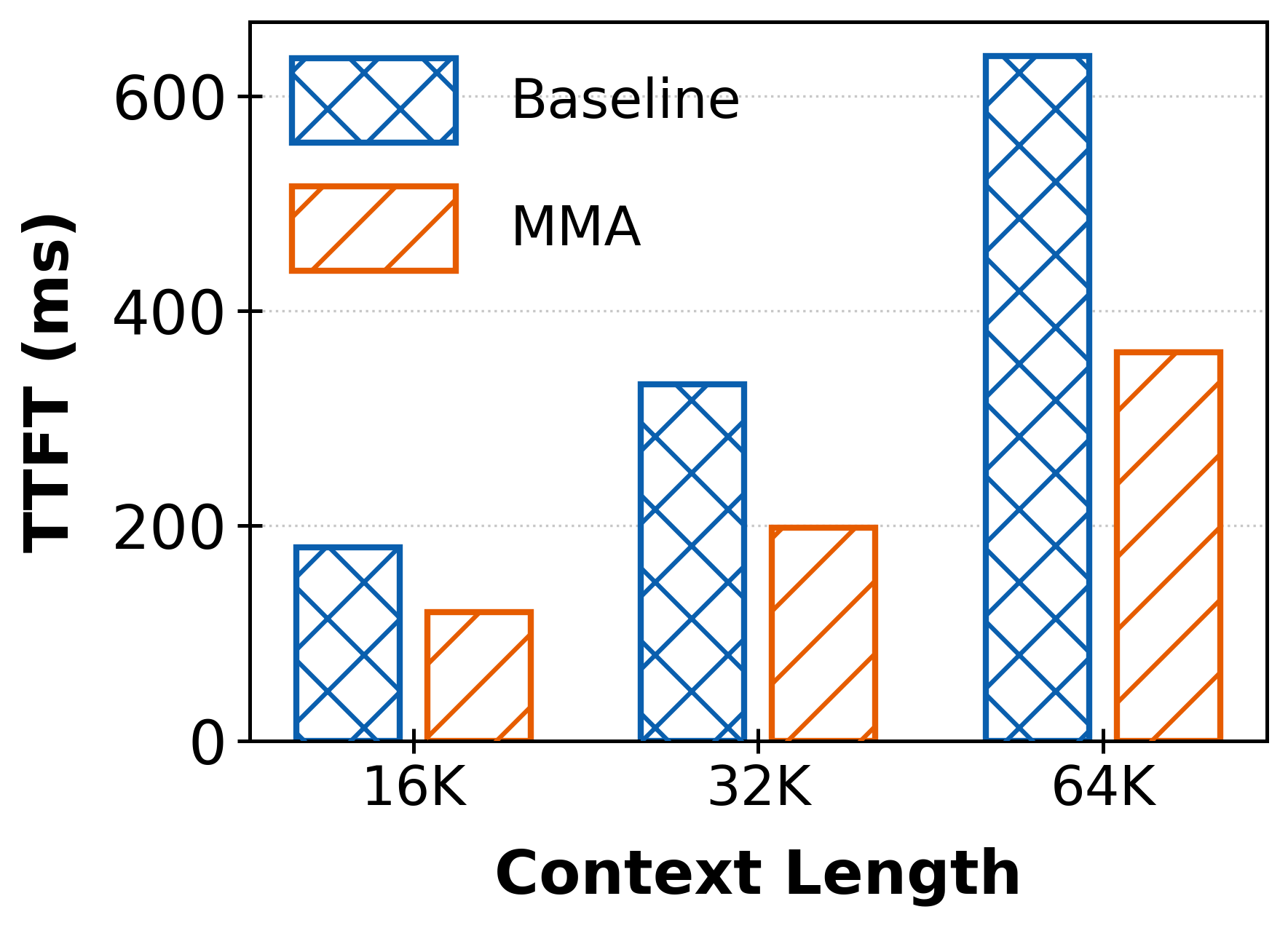}%
    \label{subfig:ttft_d}
}
\vspace{-8pt}
\caption{TTFT comparison (Baseline without MMA vs.\ MMA) under different models and context lengths.}
\label{fig:prefixcacheendtoend}
\vspace{-5pt}
\end{figure*}

\begin{figure}[ht]
    \centering
    \begin{minipage}{0.48\columnwidth}
        \centering
        \captionsetup{skip=5pt}
        \includegraphics[width=\linewidth]{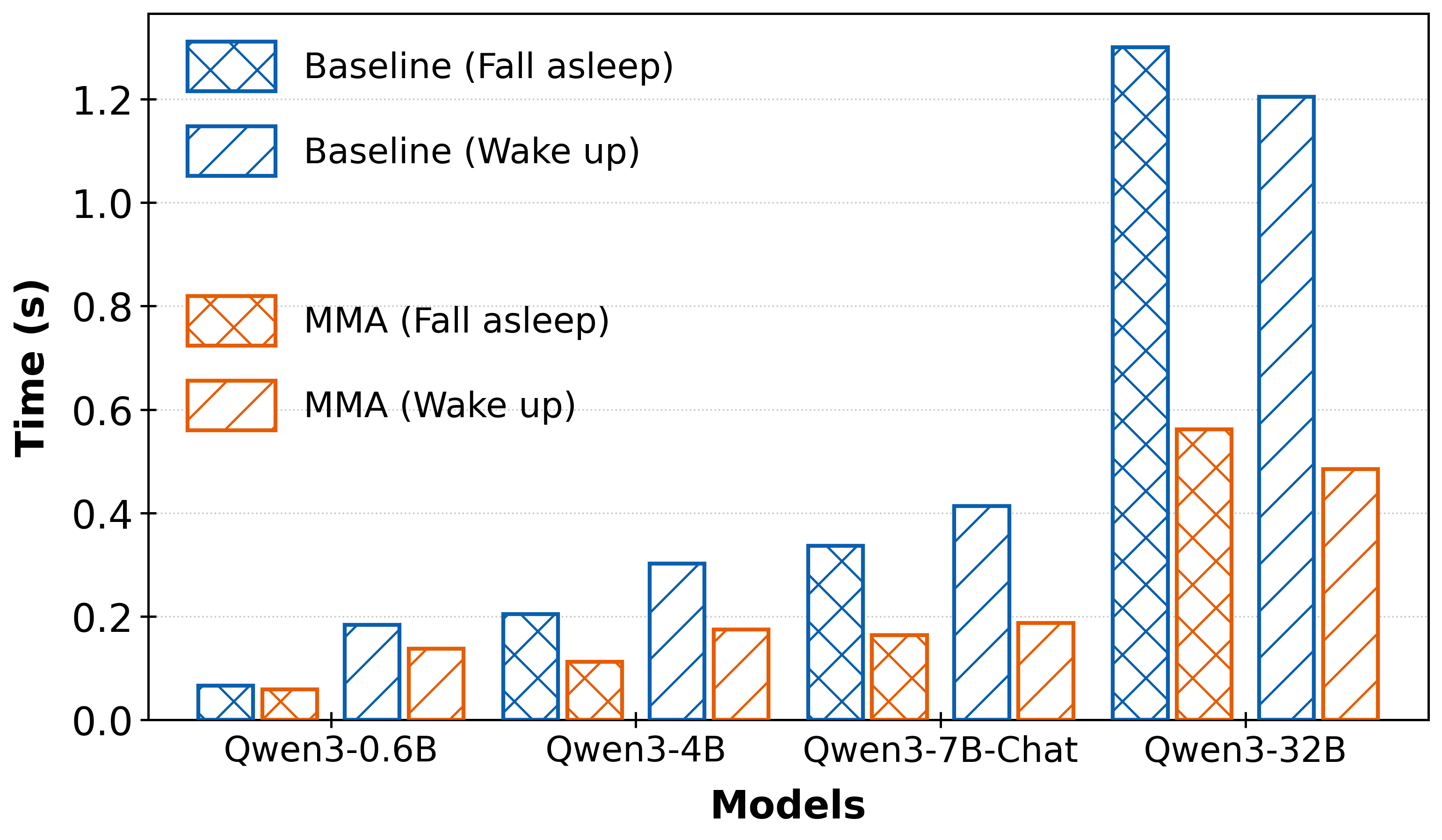}
        \captionof{figure}{Fall-asleep and wake-up time comparison.}
        \label{fig:modelswapendtoend}
    \end{minipage}
    \hfill
    \begin{minipage}{0.48\columnwidth}
        \centering
        \captionsetup{skip=5pt}
        \includegraphics[width=\linewidth]{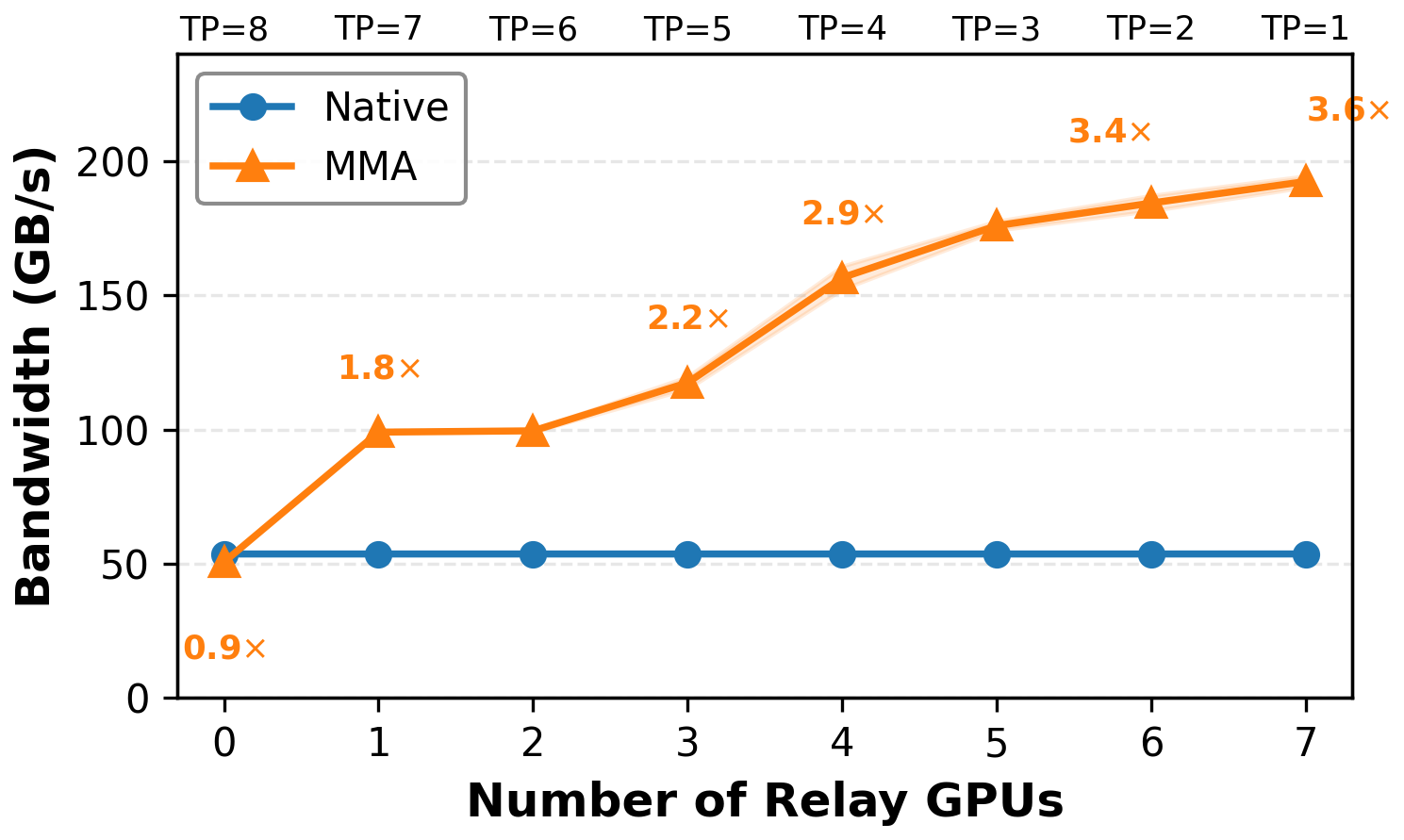}
        \captionof{figure}{MMA bandwidth vs.\ relay count.}
        \label{fig:relay-scalability}
    \end{minipage}
    \vspace{-10pt}
\end{figure}

\subsection{End-to-End Evaluation}
\label{sec:evaluation-llm-serving}

In this section, we evaluate our system in two workloads of realistic LLM serving scenarios:
(1) KV cache fetching, where cached KV pages must be transferred from host memory to GPU memory before decoding;
(2) model switching under sleep/ wake, where model weights need to be moved between CPU and GPU.

We report end-to-end improvements on TTFT and model switching latency.

\subsubsection{KV cache fetching}
\label{sec:evaluation-prefix}

\textbf{Setup.}
We evaluate how MMA affects TTFT under prefix-cache hits in the LMCache+vLLM~\cite{lmcache} inference framework with prefill--decode disaggregation.
We use four LLMs with different architectures (and therefore different KV cache sizes at the same sequence length):
Qwen3-0.6B, Qwen3-4B, Qwen-7B-Chat, and Qwen3-32B~\cite{bai2023qwen,yang2025qwen3}.
For each model, we use the LongBench v2~\cite{longbench} and select long-text documents whose context lengths are around 16K, 32K, and 64K tokens.
For each length, we take multiple documents and run multi-turn QA; we discard the first turn because it does not hit the prefix cache and report the average TTFT over the remaining turns that do hit the prefix cache.
For the 64K case, we extend max\_position\_embeddings in the local config only to ensure the model can accept 64K tokens and we do not care about generation quality in this scenario.
We report the TTFT as the main metric.

\noindent \textbf{Results.}
As shown in Figure~\ref{fig:prefixcacheendtoend}, MMA consistently reduces TTFT across all four models and all prefix lengths, achieving 1.14--2.38$\times$ speedup over the baseline.
Longer prefixes yield larger improvements because more KV cache data must be transferred and MMA's multipath advantage becomes more pronounced.

For the largest case (Qwen-7B-Chat, 64K context, 17.5\,GB KV cache), TTFT is dominated by transfer time rather than computation, and MMA delivers a 2.38$\times$ speedup.

\subsubsection{Model switching with sleep mode}
\label{sec:evaluation-sleep}

\textbf{Setup.}
We test model switching latency under vLLM's Sleep Mode~\cite{sleepmode}.
In this mode, an idle model is evicted from GPU memory (\emph{fall asleep}) and later reloaded (\emph{wake up}) when requests for that model arrive again.
Both phases mainly consist of moving model weights between CPU and GPU over PCIe, in \emph{both} directions (D2H during eviction and H2D during reload).

We evaluate four Qwen models with increasing sizes:
Qwen3-0.6B, Qwen3-4B, Qwen-7B-Chat, and Qwen3-32B~\cite{yang2025qwen3,bai2023qwen}.
For each model, we measure the wall-clock latency of the fall-asleep and wake-up phases under the baseline and MMA.

\noindent \textbf{Results.}
Figure~\ref{fig:modelswapendtoend} shows MMA reduces both fall asleep and wake up times substantially.
For the largest model, Qwen3-32B, MMA cut fall-asleep time by about 56.8\% and wake-up time by about 59.7\%, roughly 2.32--2.48$\times$ faster than the baseline.
Across all four models, MMA achieves roughly 1.12--2.48$\times$ speedup in model switching latency over the baseline.

Without changing any upper-layer scheduling policies, better use of PCIe bandwidth alone cuts large-model switching time by more than half, providing substantially more headroom to maintain TTFT SLOs under dynamic workloads.

\subsection{Sensitivity Analysis and Deep Dive} \label{Sec: Deep Dive}

\noindent \textbf{Direct priority prevents unnecessary NVLink traffic.}
To verify that direct priority avoids unnecessary NVLink contention, we launched eight concurrent H2D transfers (one per GPU, each transferring 1\,GB) with MMA enabled.
As shown in Table~\ref{tab:p2p_bandwidth}, MMA's GPU-to-GPU P2P bandwidth is essentially identical to the baseline without any concurrent transfers, confirming negligible NVLink interference.
Disabling direct priority (``MMA without direct priority'') reduces P2P bandwidth by approximately 30\,GB/s, because relay GPUs accept forwarded tasks that consume their local PCIe bandwidth and generate unnecessary NVLink traffic.

\begin{table}[htbp]
  \centering
  \caption{Influence of Direct Priority on P2P Bandwidth}
  \label{tab:p2p_bandwidth}
  \begin{tabular*}{\columnwidth}{@{\extracolsep{\fill}} l S[table-format=3.2] @{}}
    \toprule
    Method & {GPU P2P Bandwidth (GB/s)} \\
    \midrule
    P2P\_alone & 367.60 \\
    MMA & 367.28 \\
    MMA without direct priority & 330.56 \\
    \bottomrule
  \end{tabular*}
\end{table}

\begin{figure}[htbp]
    \centering
    \subfloat[Chunk Size]{
        \includegraphics[width=0.48\columnwidth]{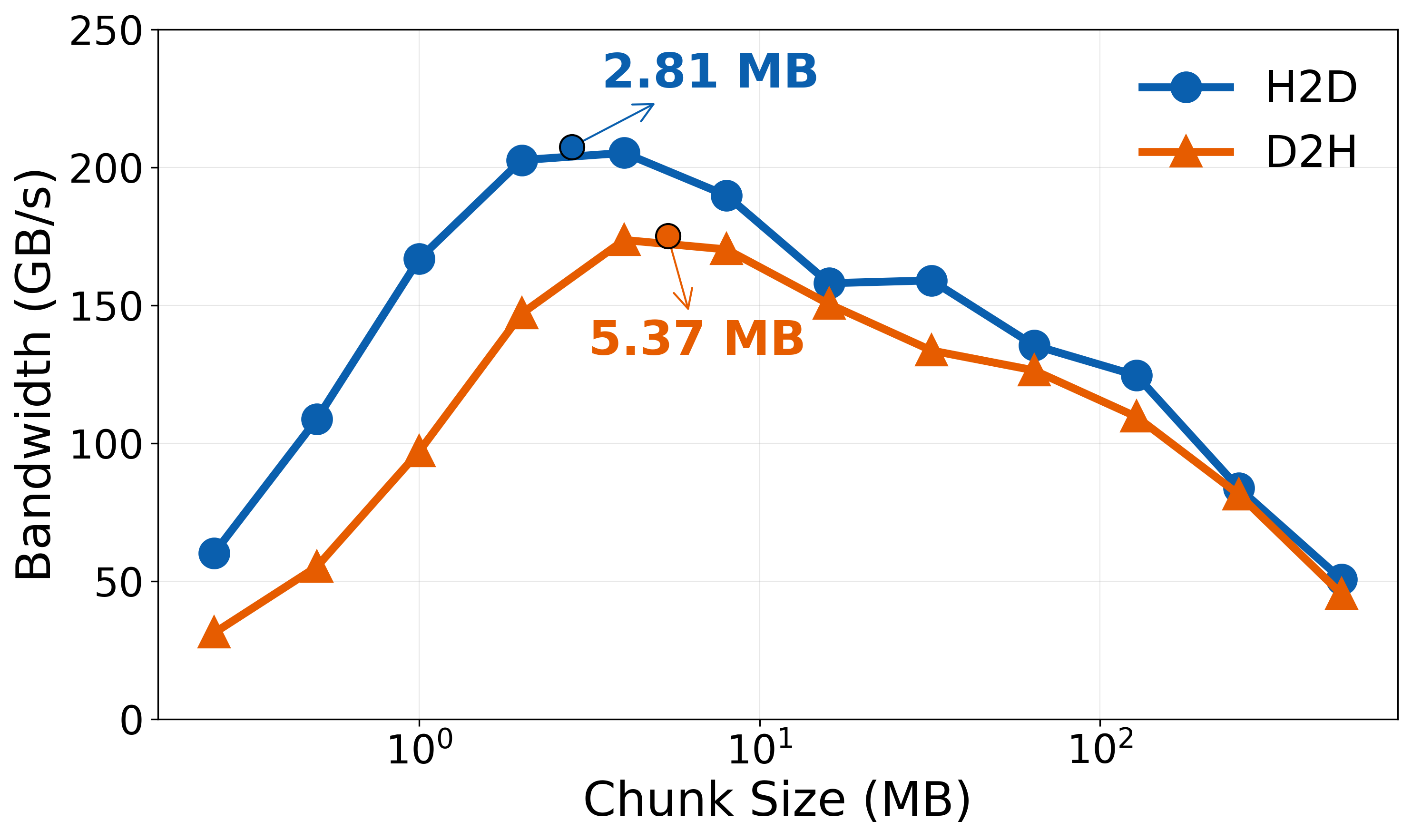}%
        \label{subfig:chunk_size}%
    }
    \subfloat[Queue Length]{%
        \includegraphics[width=0.48\columnwidth]{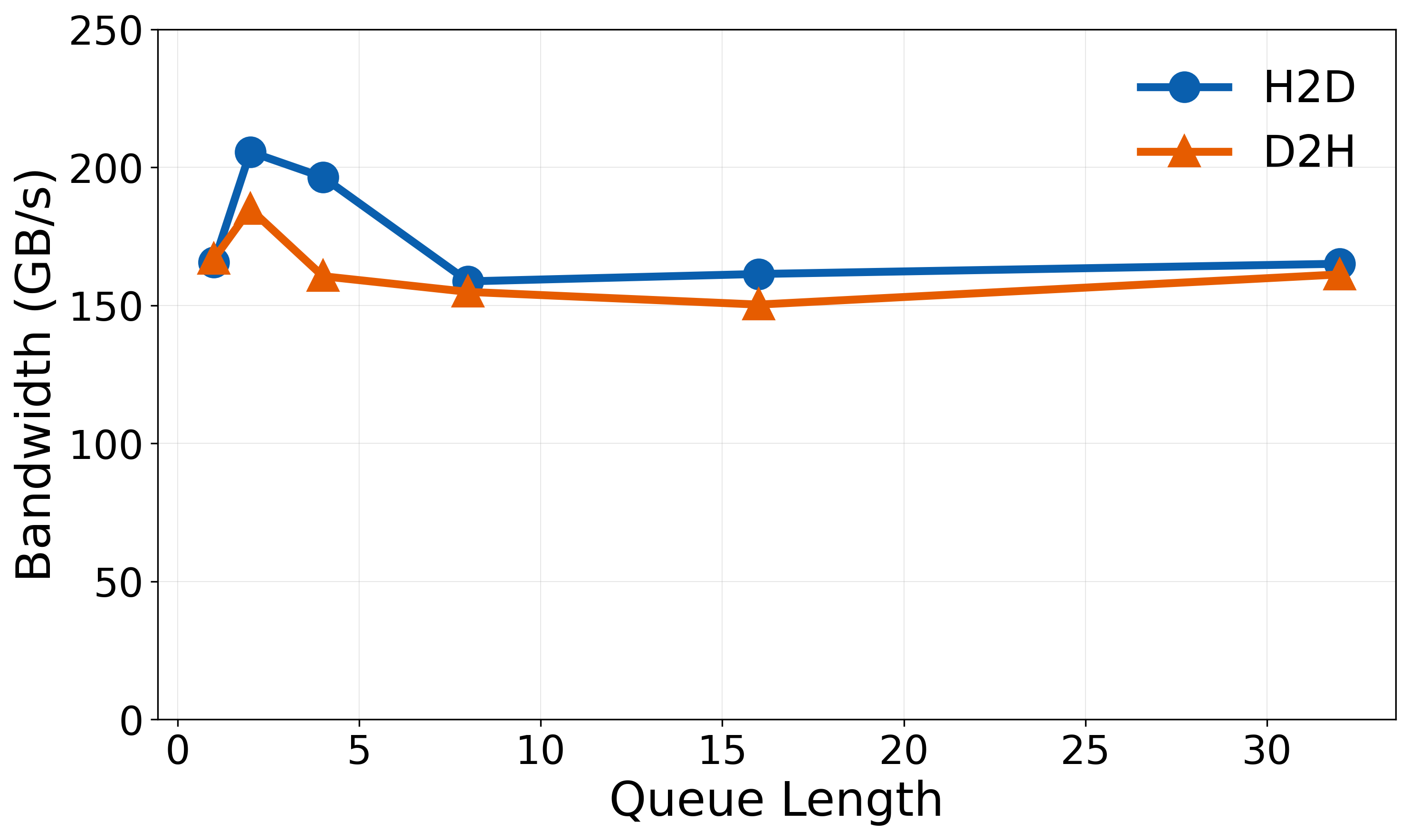}%
        \label{subfig:queue_length}%
    }
\vspace{-8pt}
\caption{Sensitivity of MMA bandwidth to chunk size and outstanding queue length.}
\vspace{-10pt}
\label{fig:chunk size and queue length}
\end{figure}

\noindent \textbf{Impact of chunk size and outstanding queue length.}
We measure how two runtime tunables---chunk size and outstanding queue length---affect MMA's transfer bandwidth on a 512\,MB workload. As shown in Figure~\ref{fig:chunk size and queue length}, H2D bandwidth peaks at a chunk size of approximately 2.81\,MB, while D2H peaks at around 5.37\,MB. An outstanding queue length of 2 yields the best bandwidth for both directions in this sweep.

Chunks that are too small incur per-task overhead, reducing GPU DMA efficiency; chunks that are too large reduce scheduling granularity and impair load balancing.
Similarly, an outstanding queue length greater than 2 coarsens scheduling granularity, while a length of~1 introduces idle gaps between consecutive transfers.
A length of~2 allows the next chunk to be enqueued before the current one completes, enabling efficient pipelining.

\begin{figure}[ht]
    \centering
    \subfloat[H2D transfer bandwidth]{
        \includegraphics[width=0.48\columnwidth]{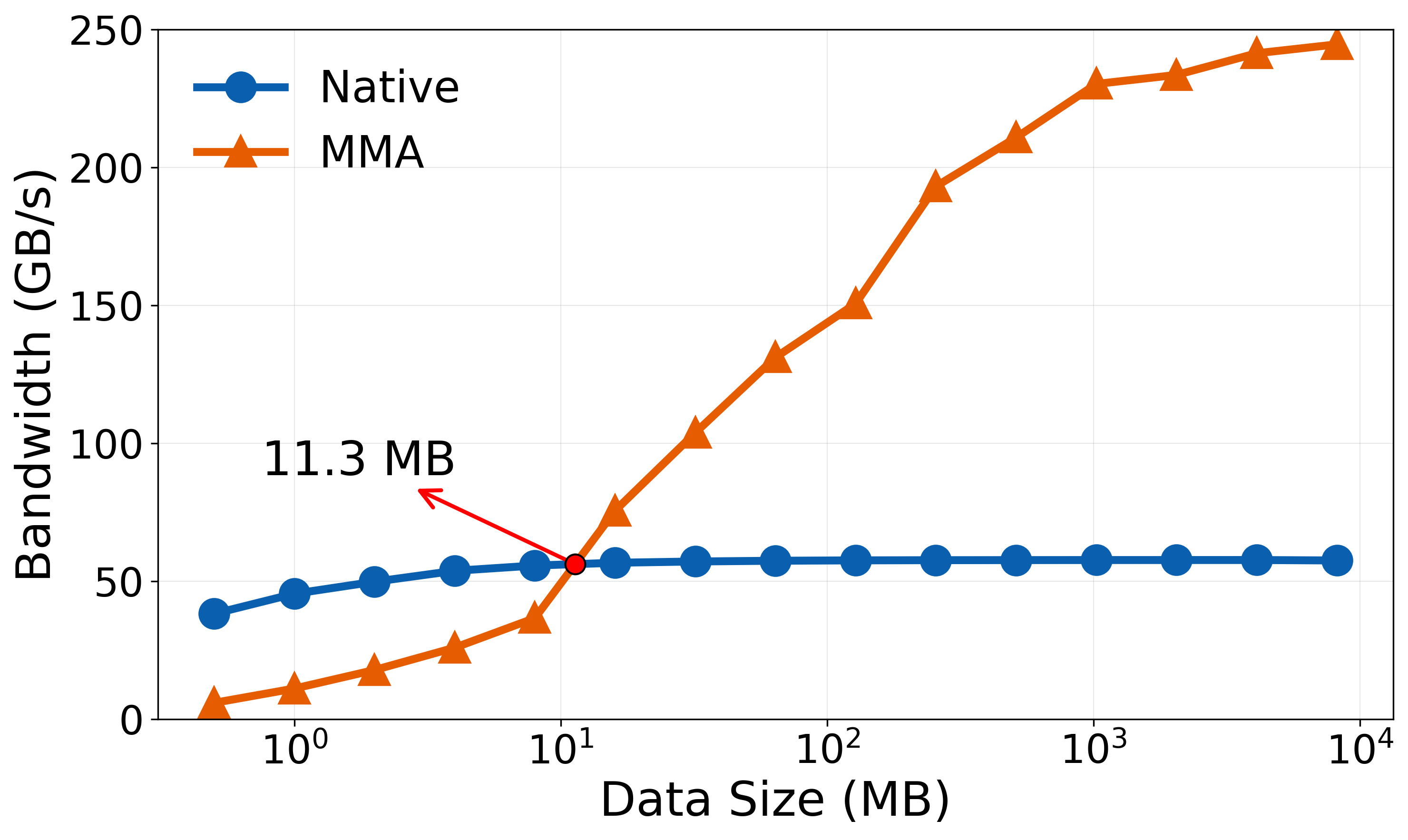}%
        \label{subfig:h2d_threshold}%
    }
    \subfloat[D2H transfer bandwidth]{%
        \includegraphics[width=0.48\columnwidth]{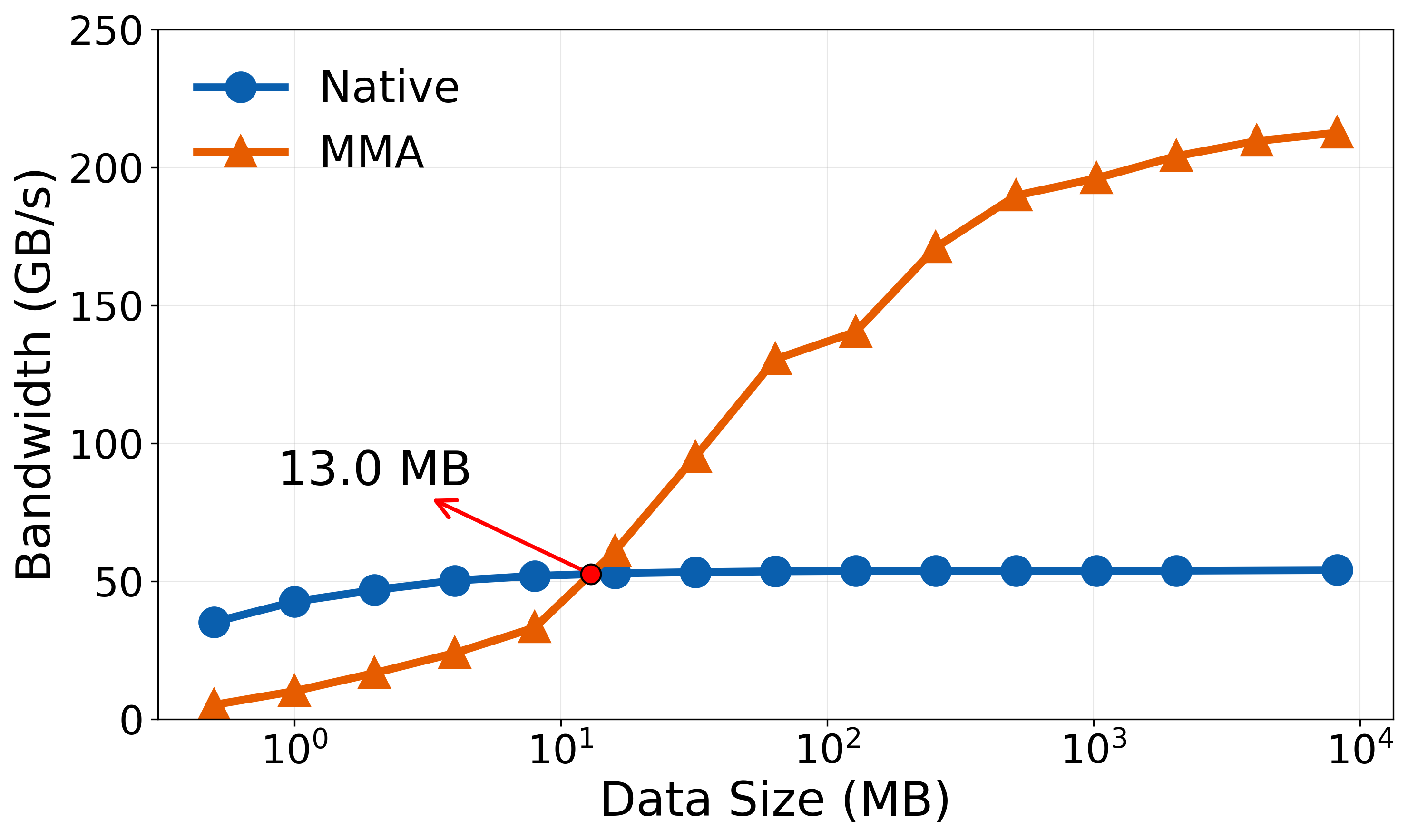}%
        \label{subfig:d2h_threshold}%
    }
\vspace{-8pt}
\caption{Optimal threshold of MMA fallback.}
\vspace{-8pt}
\label{fig:bandwidth-threshold}
\end{figure}

\noindent \textbf{Optimal fallback threshold.}
We determine the transfer size below which MMA should fall back to native single-path copy. Using the 5\,MB chunk size selected for this threshold experiment, the break-even thresholds are 11.3\,MB for H2D and 13\,MB for D2H (Figure~\ref{fig:bandwidth-threshold}). The optimal threshold falls between two and five chunks, depending on MMA's setup overhead and the effective internal transfer bandwidth.

\noindent \textbf{MMA CPU Overhead.}
In the default flow-control mode used in our experiments, each engine (H2D and D2H) spawns three threads per GPU: a transfer thread, a synchronization thread, and a lightweight monitor thread.
With both engines serving all 8 GPUs, MMA creates 48 worker threads in total.
Of these, only the 16 synchronization threads perform busy-waiting (\texttt{cudaEventSynchronize} with spin scheduling); the remaining 32 threads spend most of their time blocked on condition-variable waits or periodic sleeps and consume negligible CPU time.
Figure~\ref{fig:cpu_load_comparison} shows the additional CPU consumption introduced by MMA, expressed as equivalent fully loaded logical cores.
The overhead scales linearly with the number of active relay GPUs and reaches only 8.2 cores at 8 GPUs out of 384 logical cores available on our dual-socket EPYC~9654 platform.
In practice, during GPU-dominated LLM inference the CPU is lightly loaded (primarily running tokenization and request scheduling), so MMA's threads coexist without contention.

%% ============================================================
%% SECTION 7: Discussion and Limitations
%% ============================================================
\section{Discussion and Limitations}\label{sec:discussion}

We discuss the application scope, generality, potential and current limitations of MMA.

\noindent\textbf{Scope and sweet spot.}
MMA's benefit is proportional to the relay capacity available to a target transfer.
To quantify this, we measure the H2D bandwidth of a single target GPU while varying the number of relay GPUs from 0 to 7, emulating different tensor-parallelism (TP) configurations and different levels of peer-path availability on an 8-GPU server (Figure~\ref{fig:relay-scalability}).

At TP\,=\,1 (7 relay GPUs), MMA achieves 192.5\,GB/s---a 3.59$\times$ speedup over the 53.6\,GB/s single-PCIe baseline.
Even at TP\,=\,4, four relay paths still deliver 156.6\,GB/s (2.92$\times$).
The benefit degrades gracefully as fewer relay paths are available and shared host-side resources such as PCIe switches, host DRAM, and inter-socket links approach their aggregate limits.
At TP\,=\,8, all GPUs participate in the serving group, leaving no spare peer GPU for relay; MMA falls back to its direct path, matching the native baseline (0.94$\times$)---the slight overhead comes from MMA's chunked scheduling, which is negligible for practical purposes.

MMA is therefore best suited to transfer phases where a large object must be moved to or from a small set of target GPUs while some peer paths remain spare.
Cold-start loading and model wake-up are natural examples because they are single-target or few-target operations with large H2D volume.
At TP\,=\,8, the bottleneck shifts to host DRAM bandwidth and every GPU already has its own PCIe link, so multipath relay provides minimal additional benefit---but the per-GPU transfer volume also shrinks proportionally.
Beyond cold-start scenarios, MMA is applicable when multiple data-parallel groups are colocated on a single server, or when prefill--decode disaggregation is combined with tensor parallelism as in DistServe~\cite{distserve}, where PCIe traffic can become asymmetric across groups.
When latency predictability matters more than peak throughput, restricting relay to same-NUMA GPUs avoids the xGMI3 bottleneck entirely: four local relay paths deliver $\sim$180\,GB/s (a 3.4$\times$ improvement) with lower variance, since all traffic stays within a single memory domain.

\noindent\textbf{Relay interference on peer GPUs.}
MMA routes data through peer GPUs, raising the question of whether relay traffic degrades the peer's primary workload.
Three properties limit interference.
First, all data movement uses the GPU's DMA copy engines rather than its SMs, so compute kernels on the relay GPU are largely isolated from relay traffic at the SM level.
Second, MMA's synchronization mechanism launches a lightweight spin kernel (\texttt{<<<1,\,1>>>}) on the \emph{target} GPU's user stream; its SM footprint is negligible compared with the target GPU's compute capacity, and \texttt{\_\_nanosleep} minimizes memory-bus pressure.
Third, relay consumes NVLink bandwidth, but modestly: at MMA's peak throughput ($\sim$245\,GB/s with seven relay GPUs), each relay carries only $\sim$35\,GB/s, far below H20's per-GPU NVLink bandwidth (Table~\ref{tab:performance}).
Our robustness experiments (\S\ref{sec:design-eval}) show that when a peer GPU runs a concurrent native transfer, MMA's aggregate bandwidth degrades gracefully rather than collapsing.

\noindent\textbf{Relationship to integrated CPU--GPU architectures.}
Recent NVIDIA platforms such as Grace Hopper (GH200) and Grace Blackwell (GB200) integrate CPU and GPU on the same package via NVLink-C2C, providing up to 900\,GB/s bidirectional coherent bandwidth without PCIe involvement.
In this regime, the single-link bottleneck that MMA addresses largely disappears.
However, PCIe-based discrete GPU servers (A100, H100, H200, H20) remain widely deployed and will continue to be used for years; the PCIe bottleneck is a pressing problem \emph{today}.
Moreover, MMA's core insight---exploiting spare interconnect capacity across multiple devices to amplify a single transfer---remains valid in future architectures where analogous utilization gaps will arise across C2C or rack-scale links.

\noindent\textbf{Current limitations.}
MMA's control plane is CPU-driven, which works well for large bulk transfers but introduces non-negligible scheduling overhead for fine-grained transfers; offloading control logic to the GPU~\cite{hwang2023ark} or leveraging CUDA~12.8's batched copy interface could reduce this cost.
MMA also depends on available relay capacity: when all peer GPUs, PCIe links, NVLink ports, or copy engines are fully occupied by useful work, MMA falls back to the native direct path and provides little or no speedup.
This limitation is inherent to software-only multipath transfer rather than specific to our implementation.
MMA's benefits also depend on NUMA topology: cross-socket relay paths may become limited by inter-socket bandwidth, so topology-aware relay selection is important for predictable latency.
The current library operates within a single process: when multiple vLLM workers invoke MMA simultaneously, they may contend for the same relay GPUs.
A lightweight shared-memory daemon that arbitrates relay assignments across processes would address this; we leave it to future work.
Finally, our end-to-end evaluation focuses on individual cold-start and model-switch events; evaluating MMA under sustained, trace-driven serving workloads is an important next step.

%% ============================================================
%% SECTION 8: Related Work
%% ============================================================
\section{Related Work}

\noindent \textbf{CPU--GPU communication bottlenecks.}
Most prior work~\cite{overlapping,lmcache,prior1,flexGen,prior} operates over a \emph{single} PCIe link per GPU, relying on techniques such as pipelined transfers, double buffering, and scheduling optimizations to mitigate the bandwidth bottleneck.
These approaches aim to make one link work better rather than coordinate all available links.
In contrast, MMA expands a single host--GPU copy to use multiple available direct and relay paths when they can make progress.
To the best of our knowledge, MMA is the first system to apply software-defined multipath scheduling to host--GPU memory copies within a single server and empirically demonstrate improved effective bandwidth.
ServerlessLLM~\cite{ServerlessLLM} also employs multi-GPU parallelism to accelerate model loading: it partitions model weights across all GPUs and loads each partition through its local PCIe link simultaneously.
However, ServerlessLLM's approach requires the model to be \emph{tensor-parallelized} across all participating GPUs---each GPU loads and retains only its own shard.
MMA, by contrast, uses peer GPUs purely as \emph{transient relays}: data is forwarded to a single target GPU via NVLink after the relay GPU reads it from host memory, leaving the relay GPU's memory free for other models.
This distinction has practical implications: ServerlessLLM's approach ties GPU count to parallelism degree, while MMA decouples relay from serving---a single-GPU model can opportunistically use spare peer GPUs as relays without changing the serving configuration.

\noindent \textbf{GPU communication and inter-server multipath.}
In distributed training and inference, a large body of work~\cite{multi4,multi3,multipath,multi2,sojoodi2024multipath,temucin2021multipath} applies multipath techniques to GPU--GPU communication and inter-server bandwidth, accelerating collective operations such as AllReduce and AllToAll.
FuseLink~\cite{ren2025fuselink}, for example, jointly exploits multiple NICs and intra-server GPU interconnects, using GPUs as relays to reroute traffic between NICs and improve cross-server bandwidth and NIC utilization.
The key difference is that FuseLink builds on GPUDirect RDMA's memory registration and metadata exchange mechanisms to implement implicit GPU relay and load-aware path selection, whereas MMA targets host-to-device data transfers within a single machine, where CUDA's host--device copy abstraction provides no RDMA-like registration, metadata exchange, or network feedback channel for path control.
MMA therefore orchestrates GPU relay explicitly at the CUDA runtime layer and infers path load from local queue backpressure.
The two systems are complementary at different layers and can be combined to further improve overall communication performance.

\noindent \textbf{Multipath network protocols.}
At the data center network layer, MP-TCP~\cite{wischik2011design} and MP-RDMA~\cite{MPRDMA} have already demonstrated the effectiveness of multipath load balancing in large-scale clusters.
Unlike these protocols, which operate at the inter-host network layer and can leverage congestion signals such as RTT and ECN for load balancing, intra-server CPU--GPU DMA lacks explicit feedback and can only rely on local signals.
In addition, MMA brings multipath scheduling into the intra-server setting by treating PCIe host--GPU links and GPU interconnects such as NVLink as components of schedulable host--GPU transfer paths, providing a starting point for unified link scheduling across heterogeneous interconnects such as PCIe, NVLink, and CXL.

\noindent \textbf{Alternative data paths.}
NVIDIA GPUDirect Storage (GDS)~\cite{gds} allows data to flow directly from NVMe SSDs to GPU memory, avoiding CPU-memory staging on the data path.
GDS targets the storage$\to$GPU path and is most beneficial when models reside on local NVMe rather than host DRAM.
MMA targets the complementary DRAM$\to$GPU path; the two approaches are orthogonal and can coexist.
For the LLM serving scenarios targeted by MMA, the DRAM path remains important for two reasons.
First, KV cache offloading is inherently a DRAM operation: caches are produced by the GPU, evicted to host DRAM, and later reloaded, never touching persistent storage.
Second, even for model weights, serving stacks often cache or stage weights in host DRAM to reduce reload latency, because NVMe throughput (${\sim}$7\,GB/s per drive) is an order of magnitude below DRAM bandwidth.
CXL~3.0's fabric-attached memory~\cite{cxl3} introduces a shared memory pool accessible by both CPUs and GPUs, potentially reshaping the host--GPU transfer topology.
MMA's multipath scheduling principle---exploiting spare interconnect capacity across heterogeneous interconnects---extends naturally to CXL fabrics, where multiple CXL switches and memory controllers create additional schedulable paths.

%% ============================================================
%% SECTION 9: Conclusion
%% ============================================================
\section{Conclusion}
We have presented MMA, a software-defined multipath memory access system for host--GPU memory copies within a single server.
MMA preserves CUDA stream semantics with a dependency-preserving Dummy Task for asynchronous copies and uses a lightweight synchronization mechanism to aggregate distributed micro-transfer completion.
Through pull-based scheduling and implicit queue backpressure, MMA dispatches data across available direct and relay paths without requiring explicit link-state feedback.
Our evaluation shows that MMA achieves up to 4.62$\times$ peak bandwidth improvement and delivers 1.14--2.48$\times$ end-to-end speedups on LLM serving workloads including KV cache fetching for prefix-cache hits and model wake-up/switching.

More broadly, MMA demonstrates that spare interconnect capacity inside today's GPU servers can be harnessed through software-defined multipath scheduling without hardware, driver, or application changes.
As heterogeneous accelerators and intra-server interconnect fabrics grow more diverse (CXL, NVLink-C2C, UALink), opportunities to exploit spare intra-server I/O capacity will continue to grow.
MMA's design principles---transparent interception, stream-compatible completion aggregation, and queue-backpressure-based path selection---provide a foundation for a general intra-server multipath transport layer.
We hope this work motivates native support for exposing and scheduling multiple physical paths in future accelerator runtimes.

\bibliographystyle{ACM-Reference-Format}
\bibliography{reference}

\end{document}